\documentclass{aa}
\usepackage[pdftex]{graphicx}
\usepackage{psfig,ulem,url}
\usepackage{txfonts}
\usepackage{longtable}
\usepackage{soul,color}

\def\mathnew{\mathsurround=0pt}
\def\simov#1#2{\lower .5pt\vbox{\baselineskip0pt \lineskip-.5pt
\ialign{$\mathnew#1\hfil##\hfil$\crcr#2\crcr\sim\crcr}}}

\def\MeV{Me\kern-0.11em V}
\def\keV{ke\kern-0.11em V}

\begin{document}

\title{Galaxy clusters in the SDSS Stripe 82 \\
  based on galaxy photometric redshifts \footnote{The cluster catalogue
    will be available in electronic form at the VizieR interface of
    the Simbad
    database: http://vizier.u-strasbg.fr/viz-bin/VizieR.}}

\offprints{F. Durret  \email{durret@iap.fr}}

\author{ F.~Durret\inst{1}
\and
C. Adami\inst{2}
\and
E.~Bertin\inst{1}
\and
J.~Hao\inst{3}
\and
I.~M\'arquez\inst{4}
\and
N.~Martinet\inst{1}
\and
S.~Maurogordato\inst{5}
\and
T.~Sauvaget\inst{1,6,7}
\and
N.~Scepi\inst{1,8}
\and
A.~Takey\inst{9}
\and
M.P.~Ulmer\inst{7}
}

\institute{
UPMC-CNRS, UMR7095, Institut d'Astrophysique de Paris, 98bis Bd Arago, 
F-75014, Paris, France %1,IAP
\and
LAM, OAMP, P\^ole de l'Etoile Site Ch\^ateau-Gombert, 38 rue Fr\'ed\'eric Juliot-Curie,  13388 Marseille Cedex 13, France %2,LAM,Christophe
\and
%ETS Corporate Headquarters, 660 Rosedale Road, Princeton, NJ 08541, USA %3,Hao
Center for Particle Astrophysics, Fermi National Accelerator Laboratory, Batavia, IL 60510, USA
\and
Instituto de Astrof\'\i sica de Andaluc\'\i a, CSIC, Glorieta de la Astronom\'\i a s/n, 18008, Granada, Spain %4,Isa
\and 
OCA, Cassiop\'ee, Boulevard de l'Observatoire, B.P. 4229, F-06304 NICE Cedex 4, France %5,Sophie
\and
Observatoire de Paris-Meudon, GEPI, 92195 Meudon cedex, France %6,Tabatha
\and
Department of Physics $\&$ Astronomy, CIREA, Northwestern University, Evanston, IL 60208-2900, USA %7,Mel,Taba
\and
ENS-Cachan, 61, avenue du pr\'esident Wilson, F-94235 Cachan cedex, France %8,NicolasII
\and
National Research Institute of Astronomy and Geophysics (NRIAG), 
11421 Helwan, Cairo, Egypt %Ali
}

\date{Accepted . Received ; Draft printed: \today}

\authorrunning{Durret et al.}

\titlerunning{Galaxy clusters in the SDSS Stripe 82}

\abstract
% context heading (optional)
{The discovery of new galaxy clusters is important for two 
  reasons.  First, clusters are interesting {\it per se}, since their
  detailed analysis allows us to understand how galaxies form and evolve
  in various environments and second, they play an important part in
  cosmology because their number as a function of redshift gives
  constraints on cosmological parameters.  }
% aims heading (mandatory)
{We have searched for galaxy clusters in the Stripe~82 region of the
  Sloan Digital Sky Survey, and analysed various properties of the
  cluster galaxies.}
% methods heading (mandatory)
{ Based on a recent photometric redshift (hereafter photo$-z$) galaxy
  catalogue, we built a cluster catalogue by applying the Adami \&
  MAzure Cluster FInder (AMACFI). Extensive tests were made to
  fine-tune the AMACFI parameters and make the cluster detection as
  reliable as possible.  The same method was applied to the Millennium
  simulation to estimate our detection efficiency and the approximate
  masses of the detected clusters.  Considering all the cluster
  galaxies (i.e. within a 1~Mpc radius of the cluster to which they
  belong and with a photo$-z$ differing by less than $\pm 0.05$ from
  that of the cluster), we stacked clusters in various redshift bins
  to derive colour--magnitude diagrams and galaxy luminosity functions
  (GLFs).  For each galaxy brighter than ${\rm M_r}<-19.0$, we
  computed the disk and spheroid components by applying SExtractor,
  and by stacking clusters we determined how the disk-to-spheroid flux
  ratio varies with cluster redshift and mass.  }
% results heading (mandatory)
{We detected 3663 clusters in the redshift range $0.15\leq z \leq
  0.70$, with estimated mean masses between $\sim 10^{13}$ and a few
  10$^{14}$ M$_\odot$. We cross-matched our catalogue of candidate
  clusters with various catalogues extracted from optical and/or X-ray
  data.  The percentages of redetected clusters are at most 40\% because 
  in all cases  we detect
  relatively massive clusters, while other authors detect less massive
  structures. 
  By stacking the cluster galaxies in various redshift bins, we find a
  clear red sequence in the $(g'-r')$ versus $r'$ colour-magnitude
  diagrams, and the GLFs are typical of clusters, though with a
  possible contamination from field galaxies.  The morphological
  analysis of the cluster galaxies shows that the
  fraction of late-type to early-type galaxies shows an increase with
  redshift (particularly in 9$\sigma$ clusters) and a decrease with
  detection level, i.e. cluster mass. }
% conclusions
{From the properties of the cluster galaxies, the majority of the
  candidate clusters detected here seem to be real clusters with
  typical cluster properties.  }


\maketitle

\section{Introduction}\label{sec:intro}

The cluster count technique (e.g. Gioia et al. 1990, Allen et
al. 2011) is used to constrain cosmological parameters, and requires
catalogues with large numbers of clusters at various redshifts,
including high redshifts (z$\geq$1), and in extended fields of view
(several tens of square degrees). This is why, with the advent of
large cameras on 4m class telescopes, cluster searches at optical
wavelengths have increased in number and redshift depth over these last ten
years (see e.g. Durret et al. 2011b and references therein).

Several techniques have been applied to search for clusters, among
which we particularly want to mention the ORCA (Overdense Red-sequence
Cluster Algorithm) method, developed for the Panoramic Survey
Telescope and Rapid Response System (Pan-STARRS) described in detail
by Murphy et al. (2012) and applied by Geach et al. (2011, hereafter
GMB; see below) to the same Stripe~82 region  used in the present paper.

Other cluster searches were based on the red sequence in the colour
magnitude diagram (Erben et al. 2009, Thanjavur et al. 2009).  Among
other techniques used to search for clusters in large imaging
surveys, a matched filter detection algorithm was applied to the
Canada France Hawaii Telescope Legacy Survey (CFHTLS)
Deep fields (Olsen et al. 2007, 2008, Grove et al. 2009,
Milkeraitis et al. 2010).  The combination of optical and infrared
imaging surveys has recently led to the discovery of many high
redshift ($z>1.1$) groups and clusters (Bielby et al. 2010).  Lensing
techniques were employed to detect massive structures (i.e. with
masses larger than $10^{13}$~M$_\odot$) in the CFHTLS (e.g. Cabanac et
al. 2007, Gavazzi $\&$ Soucail 2007, Berg\'e et al. 2008, Limousin et
al. 2009). More recently, weak lensing mass measurements were made for
clusters in part of the CFHTLS Wide survey (Shan et al. 2012).  A
Bayesian cluster finder has been applied to detect galaxy clusters in
the CFHTLS by Ascaso et al. (2012) and in the Deep Lens Survey by
Ascaso et al. (2014).  Van Breukelen \& Clewley (2009) developed yet
another algorithm, named 2TecX, to search for high redshift clusters
in optical/infrared imaging surveys.  This method is based on
photometric redshifts (hereafter photo$-z$s) estimated from the full
redshift probability function and on the identification of cluster
candidates by cross-checking two different selection techniques
(adaptations of the Voronoi tessellations and of the
friends-of-friends method).  The most recent technique, redMapper, has been
developed by Rykoff et al. (2014) and applied to the SDSS DR8.

Geach et al. (2011) have searched for clusters in  Stripe~82, a
region of the Sloan Digital Sky Survey (SDSS) covering a surface of
270~deg$^2$ across the celestial equator in the Southern Galactic Cap
($-50^\circ < \alpha < 59^\circ$, $|\delta| \leq 1.25^\circ$). They
found 4098 clusters up to redshift $z\sim 0.6$ with a median redshift
z=0.32. To do this, they applied an algorithm that searches for
statistically significant overdensities of galaxies in a Voronoi
tessellation of the projected sky. They define a cluster as having 
at least five galaxy members, so we expect them to detect a higher number of
clusters than that obtained with our method. Geach et al. (2011) published a full
cluster catalogue, allowing us to compare our results directly to
theirs.

We have developed a method to search for clusters in large multiband
imaging surveys: AMACFI (Adami \& MAzure Cluster FInder, Adami \&
Mazure 1999).  We have applied it to the CFHTLS Deep and Wide fields
(Mazure et al. 2007, Adami et al. 2010, Durret et al. 2011b, hereafter
M07, A10, and D11, respectively). We have recently confirmed
spectroscopically two clusters at $z=0.61$ and $z=0.74$ detected in
the CFHTLS Deep~3 field (Adami et al. 2015a), and a third one at
  $z=0.53$ (Adami et al. 2015b), and this gives us yet more
confidence in our method.  We have also applied AMACFI to the Stripe 82 data
and present our results below.

We must keep in mind  that all these cluster searches produce
lists of cluster {\it candidates}. It is therefore important to see whether
different methods lead to the same cluster detections, and we will
therefore compare our list of cluster candidates with other available
cluster catalogues.

The paper is organized as follows. The data and method used to search for
clusters is briefly summarized in Section~\ref{sec:method}. Results on
cluster candidates are described in Section~\ref{sec:results}:
catalogue and redshift distribution.  In Section~\ref{sec:comparison}
we compare our cluster candidates to those found with other detection
algorithms.  By stacking clusters in redshift bins of 0.1, we obtained
colour-magnitude diagrams and galaxy luminosity functions, and discuss
these results in Section~5. We then compute in Section~6 the fraction
of early- to late-type galaxies in stacked clusters as a function of
redshift and of cluster mass.  A brief discussion and conclusions are
given in Section~7.

In this paper we assume H$_0$ = 70 km s$^{-1}$ Mpc$^{-1}$, $\Omega
_m$=0.3, $\Omega _{\Lambda}$=0.7.

\section{ Data and method }
\label{sec:method}

\begin{figure}
    \resizebox{\hsize}{!}{\includegraphics[width=\textwidth]{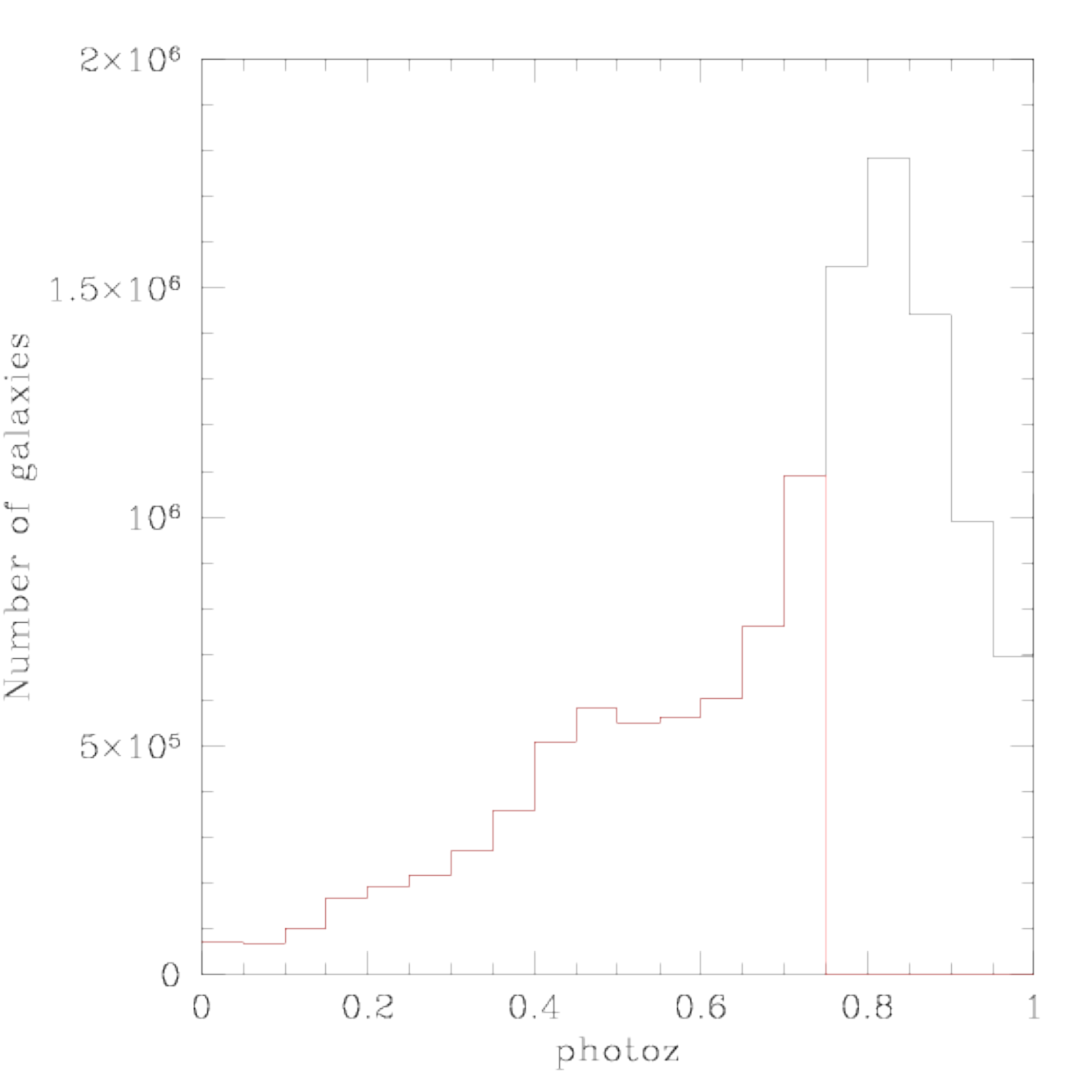}}
    \caption{Photometric redshift histogram for the initial sample of
  13,621,717 objects (black) and for the selected sample of 6,110,921
  objects with z$_{phot}\le 0.75$ (red).}
\label{fig:histo_zphot_gal}
\end{figure}

The SDSS has obtained many scans in the so-called Stripe 82 (hereafter
S82) field, defined by right ascension approximately in the range
310$^\circ-59^\circ$ and declination $|\delta| \leq 1.25^\circ$
(J2000). Five photometric bands are available: $u',\ g',\ r',\ i'$, and
$z'$.  These repeated observations have been averaged to produce deeper
and more accurate photometry than the nominal 2\% single-scan
photometric accuracy (Ivezi\'c et al.  2004).

\subsection{Stripe~82 catalogues}
\label{sec:cats}

We started with the Msplit catalogue of 13,621,717 objects  
available in the SDSS database. For each
object this catalogue contains  the SDSS identification (19 digit number), right ascension,
declination, photo$-z$, and error on the photo$-z$ made by Reis et
al. (2012), and is limited in magnitude to $r'<24.5$.  The photo$-z$
histogram of these 13,621,717 objects is shown in
Fig.~\ref{fig:histo_zphot_gal}. To avoid incompleteness (which becomes
apparent in Fig.~\ref{fig:histo_zphot_gal} for z$_{phot} \sim 0.8$),
we cut this catalogue at z$_{phot}\le 0.75$ and were then left with
6,110,921 objects. This photo$-z$ catalogue was used to detect cluster
candidates.

\begin{figure}
\centering
   \resizebox{\hsize}{!}{\includegraphics[width=\textwidth]{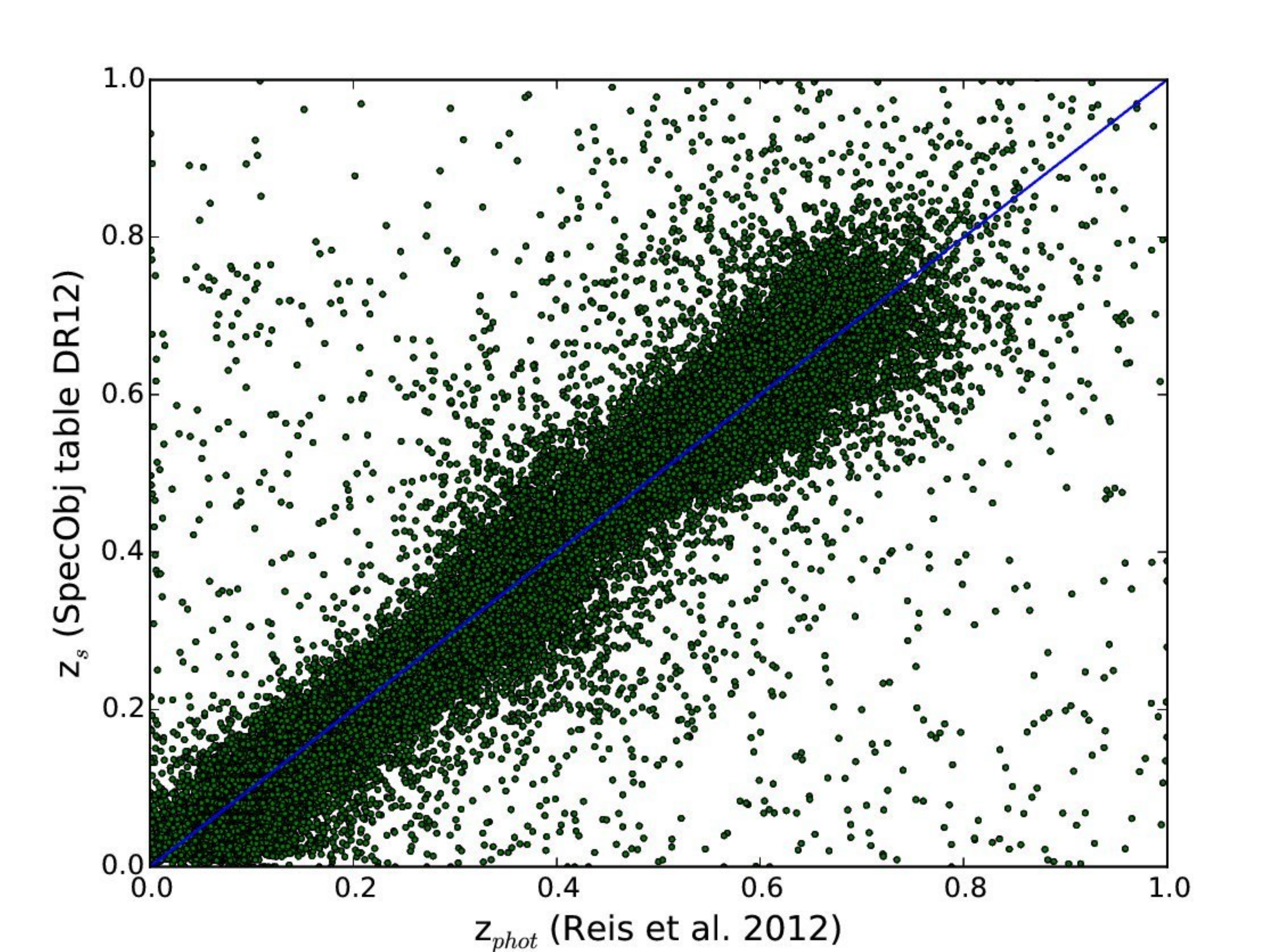}}
   \caption[]{Spectroscopic redshifts in the S82 region taken from SDSS
  DR12 versus photometric redshifts from Reis et al. (2012) for 105,613
  galaxies.  The blue line shows the diagonal of the square.}
\label{fig:zp_zs}
\end{figure}

As  a check to the quality of the Reis photo$-z$ catalogue, we
cross-correlated it with the SDSS spectroscopic catalogue, {\it
  SpecObj} table of the recent data release DR12. The result is shown
in Fig.~\ref{fig:zp_zs} for 105,613 galaxies. For the difference
$|z_{p,{\rm Reis}} - z_s|$ between the Reis photo$-z$s $z_{p,{\rm
    Reis}} $ and the spectroscopic redshifts $z_s$, the mean value is
0.027, the median is 0.016, and the standard deviation is 0.047.  As a
comparison, we made the same correlation between the DR12 photo$-z$s
extracted from the {\it Photoz} table and the spectroscopic redshifts
$|z_{p,{\rm DR12}} - z_s|$ and found a mean value of 0.038, a median
of 0.023, and a standard deviation of 0.053. This confirms that the
Reis photo$-z$ catalogue is better than the general {\it Photoz} DR12
catalogue, and we will therefore use the Reis catalogue for our
analysis. The fact that there are very few spectroscopic redshifts
above $z\sim 0.8$ to  calibrate the photo$-z$s justifies our cut at
$z_{phot}=0.75$.

We also retrieved the dereddened magnitude catalogue of 8,485,885
objects (Annis et al. 2014) which we later cross-correlated with the
photo$-z$ catalogue to obtain a complete catalogue of 4,999,968
galaxies that was fed into the Le~Phare software (Arnouts et al. 1999,
Ilbert et al. 2006) to compute the absolute magnitudes that we will
exploit further in the paper.

We first considered the curves shown in Fig.~8 of Annis et al. (2014)
to estimate the 90\% completeness limit of our magnitude catalogue.
These give the following approximate values: $u'_{lim}$=23.1,
$g'_{lim}$=22.8, $r'_{lim}$=22.4, $i'_{lim}$=22.1, and $z'_{lim}$=20.4.
However, when drawing the magnitude histograms in the five photometric
bands (see Fig.~\ref{fig:histomagall}) and superimposing these limits
(marked as dotted vertical lines), we found that although in the $u'$
and $g'$ bands the Annis limits seemed acceptable (i.e. brighter than
the magnitude when incompleteness becomes obvious), in the $r'$ and $i'$
bands these limits were obviously too faint while in the $z'$ band
the limit was too bright.  We therefore  take the following
(rather conservative) 90\% completeness limits: $u'_{lim}$=23.0,
$g'_{lim}$=22.8, $r'_{lim}$=22.1, $i'_{lim}$=21.5, and $z'_{lim}$=21.2
(marked as full vertical lines in Fig.\ref{fig:histomagall}).

\subsection{ Method for cluster detection}

\subsubsection {Overall description of the method}

We applied to this photo$-z$ catalogue the same treatment as in M07,
A10, and D11, where a full description is given.  This method has also
been applied by A10 to the Millennium simulation (Springel et
al. 2005) to assess the quality of the detections and to obtain a
rough estimate of the relation between the cluster masses and the
significance level at which clusters were detected. We have done the
same for the S82 data, as described below.

We first divided the photo$-z$ catalogue in slices of 0.1 in redshift,
each slice overlapping the previous one by 0.05 (i.e. the first slice
covers redshifts 0.1 to 0.2, the second 0.15 to 0.25, etc. and the
last slice is 0.65--0.75). As discussed by A10, the 0.1 redshift width
of the studied slices is the best compromise between the redshift
resolution and the possible dilution of the density signal due to
typical photometric redshift uncertainties. Then, to make the data
manageable (in ram-active CPU memory), each subcatalogue was then
divided into slices of 1.1~deg in right ascension, with an overlap of
0.1~deg between slices. No cut was made in declination.

We built galaxy density maps for each redshift slice, based on the
adaptative kernel technique described in M07, with a pixel size
(originally taken to be 1~arcmin) that will be discussed below and 100
bootstrap resamplings of the maps to estimate  the background
level correctly.

We then detected structures in these density maps with the SExtractor
software (Bertin \& Arnouts 1996) in the different redshift bins at
various significance levels: 3$\sigma$, 4$\sigma$, 5$\sigma$,
6$\sigma$, and 9$\sigma$ (as defined by SExtractor).

The structures were then assembled in larger structures (called
$detections$ in the following) using a minimal spanning tree
friends-of-friends algorithm (see Adami \& Mazure 1999). Two
$detections$ with centres  distant by less than 2~arcmin (twice the
pixel size defined originally) were merged into a single one which was
assigned the redshift of the $detection$ having the highest S/N.  We
did not merge $detections$ within 2~arcmin into a single one if their
photometric redshifts differed by more than 0.09 to avoid losing
clusters that could be almost aligned along the line of sight
but located at very different redshifts. With this separation limit
(hereafter called the separation parameter), the typical uncertainty
on cluster positions is therefore about 2~arcmin.  This respectively
corresponds to 310~kpc and 860~kpc at $z=0.15$, the lowest redshift,
and $z=0.7$, the highest redshift in our cluster sample. We  also
briefly discuss below the influence of the choice of this separation
limit on the final cluster catalogue.

\subsubsection{Choice of  pixel size for the density maps and of the
separation parameter}

We  initially built galaxy density maps for each redshift slice,
with a pixel size of $1.002\times 1.002$~arcmin$^2$. With this
  pixel size we obtained a cluster catalogue containing 956 clusters
  in the redshift range of 0.15--0.7.  Since S82 covers an area of
  270~deg$^2$, the spatial density of this catalogue is 956/270 =
  3.54~clusters~deg$^{-2}$, while if we consider the clusters detected
  in the CFHTLS--Wide~1, Wide~2, Wide~3, and Wide~4 (D11) in the same
  redshift range ($0.15 \leq z \leq 0.70$), we find respective
  densities of 17.0, 15.9, 14.8, and 16.6~clusters~deg$^{-2}$, using a
  pixel size of 0.54~arcmin and a separation parameter of 3~arcmin.
When the search for clusters in the CFHTLS was made, the separation
parameter was still an angle, while  in the minimal spanning
tree code we now implement a separation in Mpc, which is more physical. So
  our detection level in S82 was smaller than that of the CFHTLS--Wide
  by a factor between 4.2 and 4.8.  A first explanation could be that
  the S82 catalogue is shallower than that of the CFHTLS, and does not
  reach similar redshifts and/or magnitudes. However, if we compare
  the galaxy photo$-z$ histogram of the S82 to that of the CFHTLS Wide
  survey (Fig. 2 in D11, black line), we can see that the S82
  histogram starts decreasing for z$>$0.85, while the CFHTLS--Wide
  starts decreasing for z$>$0.90, so the photo$-z$ completeness limit
  of S82 is lower than that of the CFHTLS--Wide only by $\sim
  0.05$. If we compare the magnitude completeness limits of the two
  surveys, the S82 90\% completeness limit is reached for r$\sim22.1$
  according to Annis et al. (2014). In the CFHTLS--Wide,
  incompleteness begins to show for $i' \sim 23.5$, which corresponds
  to $r'$ between 22.5 and 23 (for an elliptical galaxy at redshift
  0.2 or 0.5, respectively), showing that the S82 catalogue is
  shallower that the CFHTLS--Wide only by approximately half a
  magnitude.  So the discrepancy by a factor of 4 between the density
  of clusters detected with AMACFI in the two surveys seems too large
  to be explained only by their difference in depth. This led us to
  question our method and to make several tests, first on the pixel
  size chosen to compute the density maps on which our cluster
  detection is based, and second on the separation parameter.

Since the CPU time necessary to compute density maps increases
  dramatically as the pixel size decreases (and hence the
  number of pixels increases), we made the following tests on a subregion of
  S82 covering $1<$RA$<10$~deg, with the full declination range
  $|\delta |\leq 1.25^\circ$. We considered pixel sizes of $30 \times
  30$~arcsec$^2$, $15 \times 15$~arcsec$^2$, and $10 \times
  10$~arcsec$^2$. As the pixel size decreases, the number of
  structures detected increases, so the completeness of the cluster
  catalogue increases. However, we must be careful not to start
  detecting very small structures that cannot be clusters, because in
  this case the purity of the cluster catalogue will decrease.

As mentioned above, we took a separation parameter of 2~Mpc.
 Since the separation parameter could have an influence
  on the number of candidate clusters detected, we made tests with
  separation parameters of 1~Mpc, 2~Mpc, and 3~Mpc, and
  the results are given below.

We also tested how the quality of the photo$-z$s could influence
  the numbers of candidate clusters detected by applying two different
  selections. First, we considered only the galaxies with an error
  $\delta_{zp} \leq 0.1$ on their photo$-z$. Such a cut reduces the
  number of galaxies with photo$-z \leq 0.75$ from 6,110,921 to
  2,458,235, and therefore excludes 59.8\% of the galaxies.  Second,
  we considered only the galaxies with a relative error smaller than
  50\%: $\delta _{zp}/z_p \leq 0.5$.  In this case, the total number of
  galaxies drops from 6,110,921 to 4,469,271, and thus we exclude
  26.9\% of the galaxies.  

\begin{figure}
\centering
   \resizebox{\hsize}{!}{\includegraphics[viewport=0  0 650 450,clip]{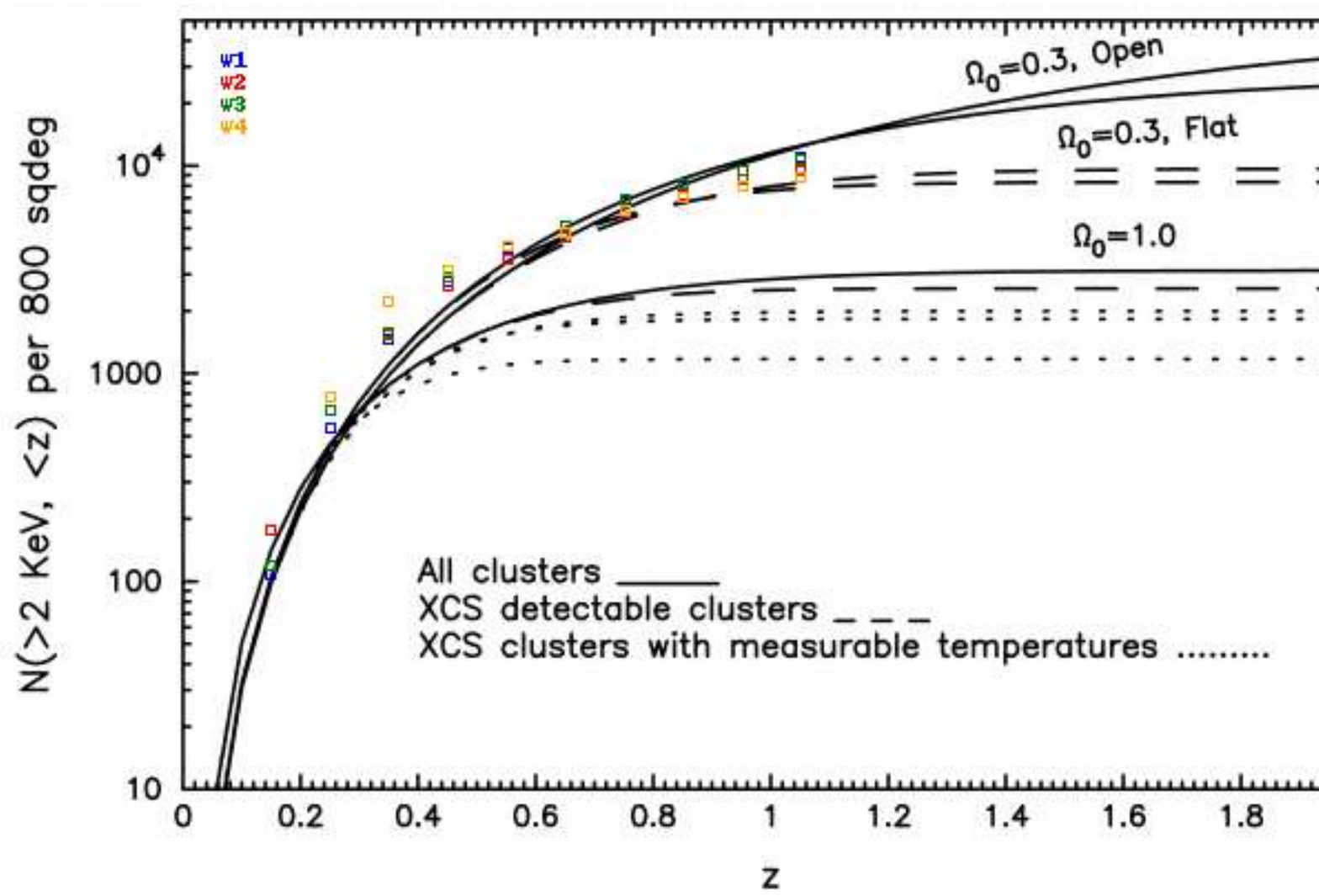}}
   \resizebox{\hsize}{!}{\includegraphics[viewport=0  0 650 450,clip]{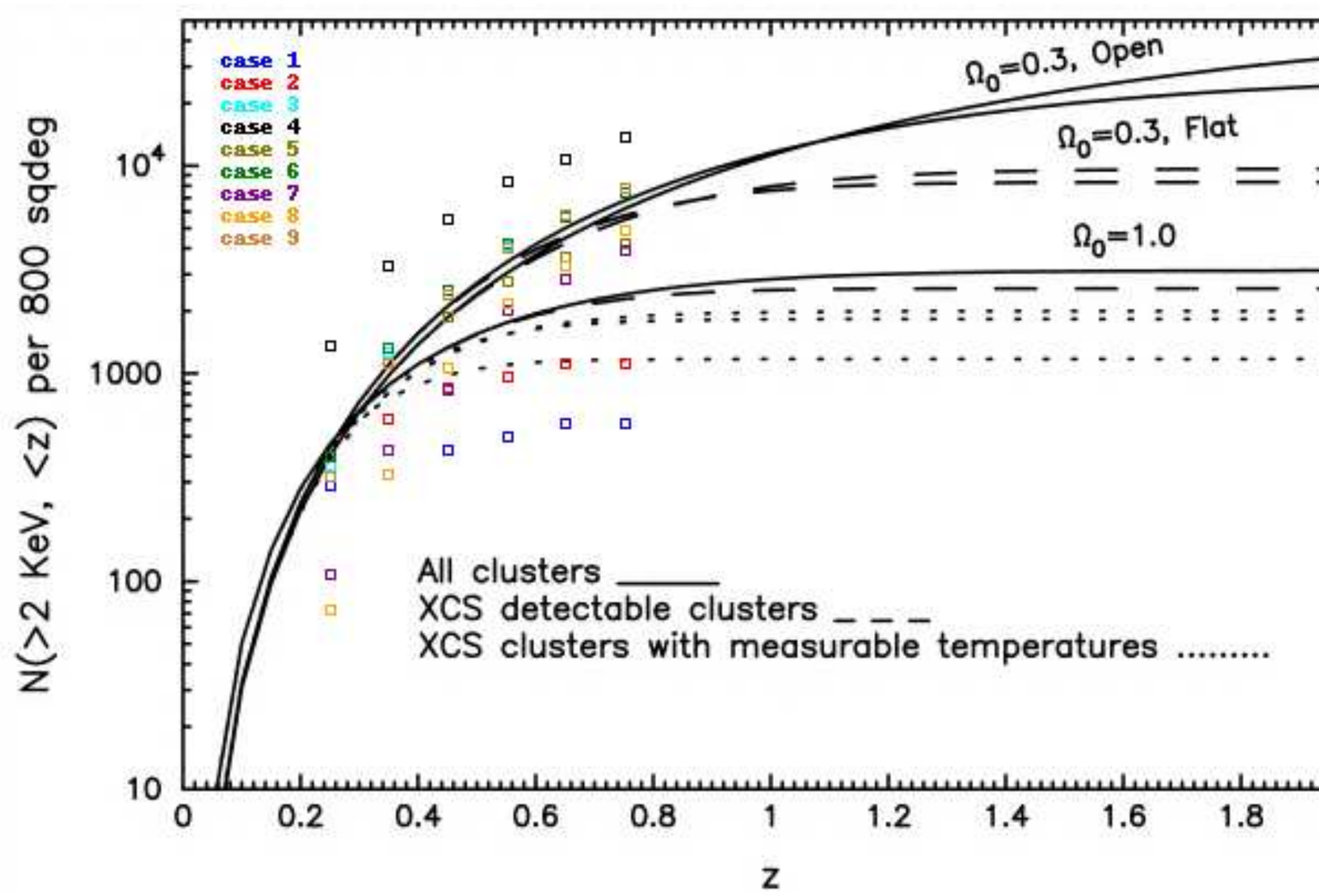}}
   \caption{Cumulative number of clusters hotter than 2~keV expected in
  a region of 800~deg$^2$ for different cosmologies as a function of
  redshift, taken from Romer et al. (2001), Fig.~5.  The numbers of
  clusters in the four CFHTLS--Wide fields are shown in the top
  figure, and the numbers detected in  Stripe~82 for the various
  cases described in Table~\ref{tab:cases} (see text) are plotted in
  the bottom figure. Only clusters detected at a 4$\sigma$ level and
  above are taken into account. }
\label{fig:Romer}
\end{figure}

In order to have an objective criterium for the choice of the
  cluster detection parameters, we considered the plots showing the
  cumulative number of clusters hotter than 2~keV expected in a region
  of 800~deg$^2$ for different cosmologies as a function of redshift,
  taken from Romer et al. (2001), Fig.~5b.  The mass-temperature
  relation of Xu et al. (2001) implies that kT$>$2~keV corresponds to
  clusters of masses M$_{r200}>(1.2-1.6)\times 10^{14}$~M$_\odot$. As
  a first test, we overplotted on these curves the densities of
  clusters detected by D11 in the four CFHTLS Wide fields. We found a
  very good agreement between our cluster densities and the Romer
  curves for $\Omega _0 =0.3$ when considering the clusters detected
  at 4$\sigma$ and above, as illustrated in Fig.~\ref{fig:Romer}
  (top).
We then overplotted on the same curves the densities of clusters
  detected at a 4$\sigma$ level in a subregion of S82 defined by
  $1<$RA$<10$~deg and $|\delta | \leq 1.25^\circ$ for the nine 
  cases summarized in Table~\ref{tab:cases}.

\begin{table}
  \caption{Cases considered in the various tests on cluster detection parameters. 
    The columns are: (1)~running number of test; (2)~pixel size chosen to compute 
    the density maps; (3)~separation, that is the minimum value above which two
    detected structures are considered to be different if they differ by
    more than 0.09 in redshift; (4)~cut depending on the photometric redshift 
    uncertainty when applicable.  }
\begin{tabular}{cccc}
\hline
\hline
Case & pixel size & separation & cut on $z_p$  \\
     & (arcsec$\times$~arcsec) & (Mpc) & \\ 
\hline
1  & 60$\times$60 & 2 & $-$ \\
2  & 30$\times$30 & 2 & $-$ \\
3  & 10$\times$10 & 2 & $-$ \\
4  & 10$\times$10 & 1 & $-$ \\
5  & 15$\times$15 & 2 & $-$ \\
6  & 10$\times$10 & 2 & $\Delta z_p/z_p \leq 0.5$ \\
7  & 10$\times$10 & 3 & $\Delta z_p/z_p \leq 0.5$ \\
8  & 10$\times$10 & 3 & $\Delta z_p \leq 0.1$ \\
9  & 10$\times$10 & 3 & $-$ \\
\hline
\end{tabular}
\label{tab:cases}
\end{table}

In  Table~\ref{tab:cases}, for each case we give the
  pixel size chosen to compute the density maps and the
  ``separation'', that is the minimum value above which two
  detected structures are considered to be different if they differ by
  more than 0.09 in redshift.  In some cases we have also applied a
  cut based on the error on the photo$-z$.

We now compare the numbers of clusters detected at a 4$\sigma$ level
and higher in   Stripe~82 to the Romer et al. (2001) curves, as
shown in Fig.~\ref{fig:Romer} (bottom). First, we can see that our
original choice of parameters (case~1) leads to a
number of clusters that is much too small. Pixel sizes of $30 \times 30$~arcsec$^2$ (case~2)
and $15 \times 15$~arcsec$^2$ (case~5) also lead to too few clusters.   If we take a separation equal to 1~Mpc or 3~Mpc
(cases 4 and 7, respectively), the numbers of clusters fall clearly
above and below the Romer curve, so we decided to keep a
separation of 2~Mpc. The number of clusters detected in case~8 is also
much too small. The best match with the Romer et al. curves is obtained
for cases~3, 6, and 9. In order to keep our sample as similar as possible
to the cluster sample extracted in the CFHTLS-W, we chose to make the
cluster detection on the full catalogue (limited to $z_p\leq 0.75,$ but
with no condition on the photo$-z$ error) with a pixel size of $10
\times 10$~arcsec$^2$ and a separation of 2~Mpc (case~3). 

In this way we obtained a final catalogue of 3663 candidate clusters
detected at a significance level from 3$\sigma$ to 9$\sigma$. This
catalogue-- including for each cluster the coordinates, photo$-z$,
detection level and number of cluster galaxies-- will be available at
the VizieR interface of the Simbad
database\footnote{http://vizier.u-strasbg.fr/viz-bin/VizieR}.

\section{Cluster catalogue}
\label{sec:results}

\subsection{Significance level and spatial distribution of the
  candidate clusters}

\begin{figure}
\centering
  \resizebox{\hsize}{!}{\includegraphics[viewport=50  160 590 710,clip]{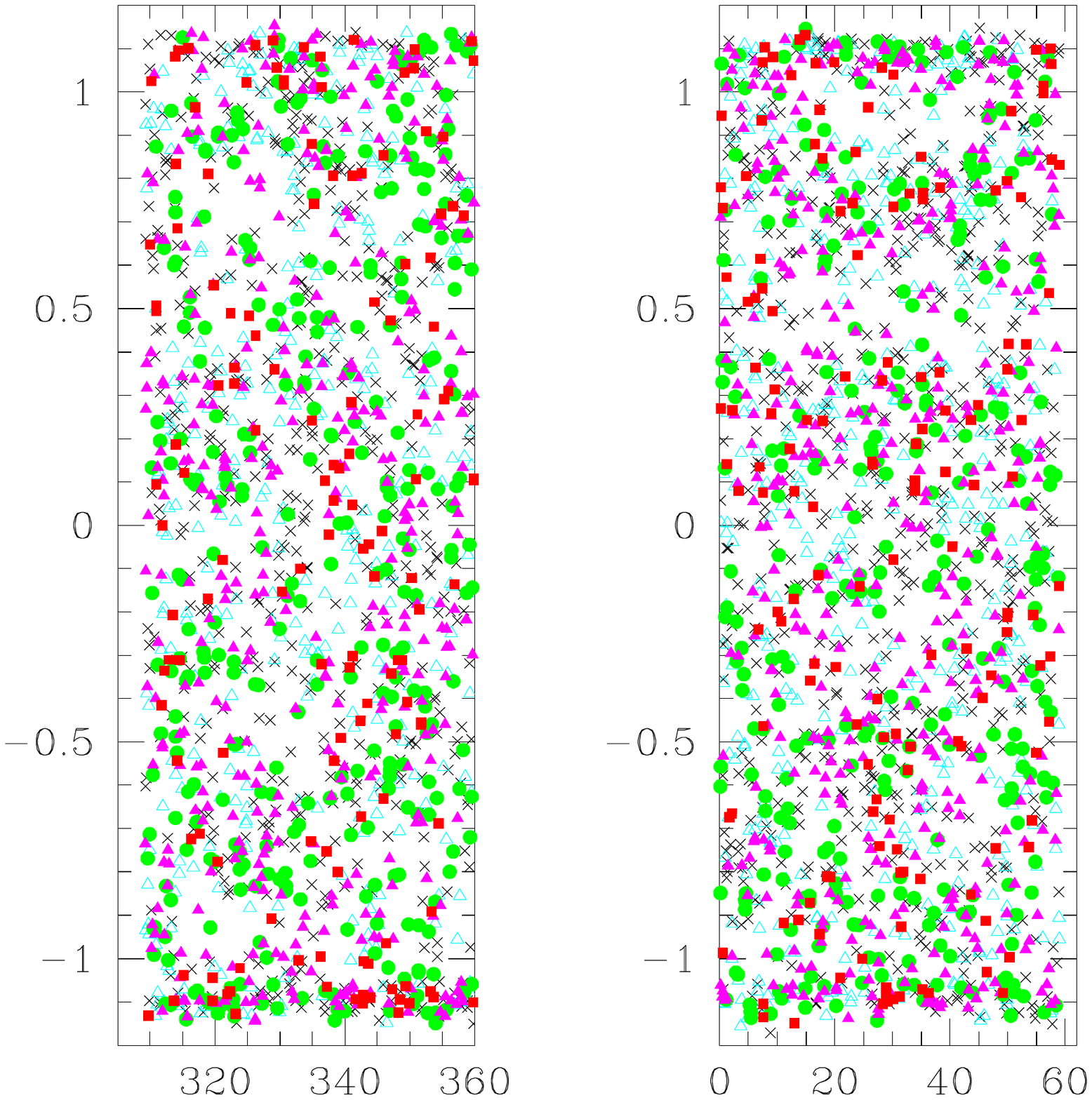}}
  \caption[]{Projected spatial distribution of the 3663 candidate
  clusters, colour coded as follows: red squares: S/N=$9\sigma$,
  magenta filled triangles:S/N$= 6\sigma$, green filled circles:
  S/N$=5\sigma$, cyan empty triangles: S/N$= 4\sigma$, black crosses:
  S/N$= 3\sigma$. }
\label{fig:xyclu}
\end{figure}

In the catalogue of 3663 cluster candidates, the numbers of clusters
detected at the various significance levels of 3$\sigma$, 4$\sigma$,
5$\sigma$, 6$\sigma$, and 9$\sigma$ are: 1133, 792, 623, 820, and 295,
respectively. In Sections 5 and 6 we  concentrate on the
properties of the 2530 clusters detected at $4\sigma$ and above to
limit our analysis to the objects that are the most likely to be real
clusters.

The projected spatial distribution of all the detected clusters is
shown in Fig.~\ref{fig:xyclu} with different symbols for the various
significance levels. We can see concentrations of candidate clusters
at the edges in declination, for $|\delta|\sim 1.1-1.2$, which are
most probably spurious detections. We keep these objects in our final
catalogue for the sake of completeness, but we note this
shortcoming.

\subsection{Redshift distribution of the candidate clusters}

\begin{figure}
\centering
\resizebox{\hsize}{!}{\includegraphics[viewport=50  160 590 710,clip]{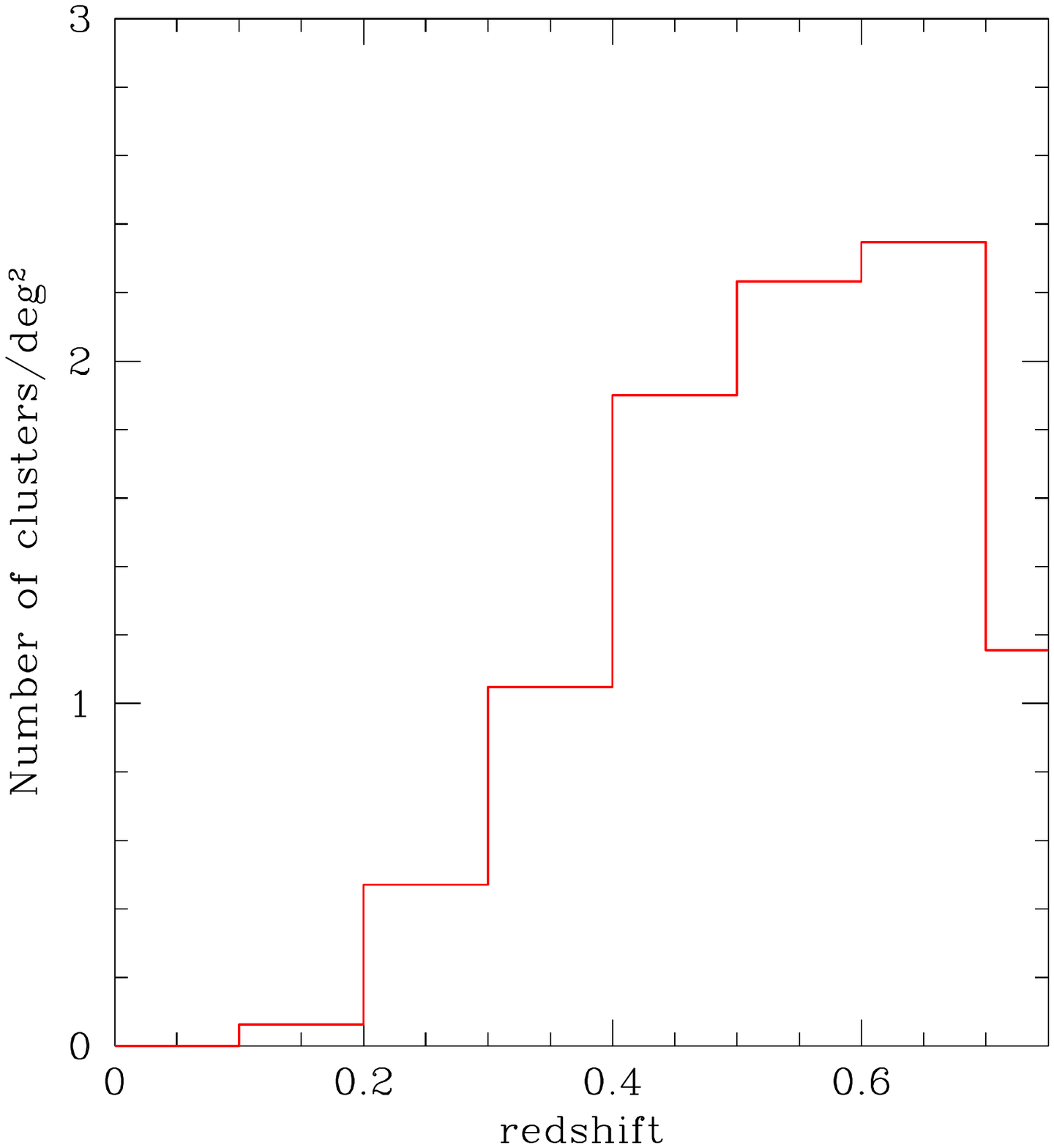}}
\caption[]{Histogram of the surface density of cluster candidates in
  S82 as a function of photometric redshift. }
\label{fig:histoz_clu}
\end{figure}

The photometric redshift distribution of the 3663 candidate clusters
detected in S82 in the redshift range 0.15--0.7 (divided by
270~deg$^2$ to obtain a surface density directly comparable to those
found in the literature) is shown in
Fig.~\ref{fig:histoz_clu}.  This photo$-z$ distribution has a
  mean value of 0.51 and a median of 0.53, with dispersions of 0.15
  around these values.  The median redshift of our clusters is notably
  higher than the median redshift z=0.32 found by Geach et al. (2011)
  for their sample of 4098 clusters, and the comparison between both
  samples will be made in Section~\ref{sec:comparison}.

\begin{figure}
\centering
  \resizebox{\hsize}{!}{\includegraphics[viewport=10  150 590 710,clip]{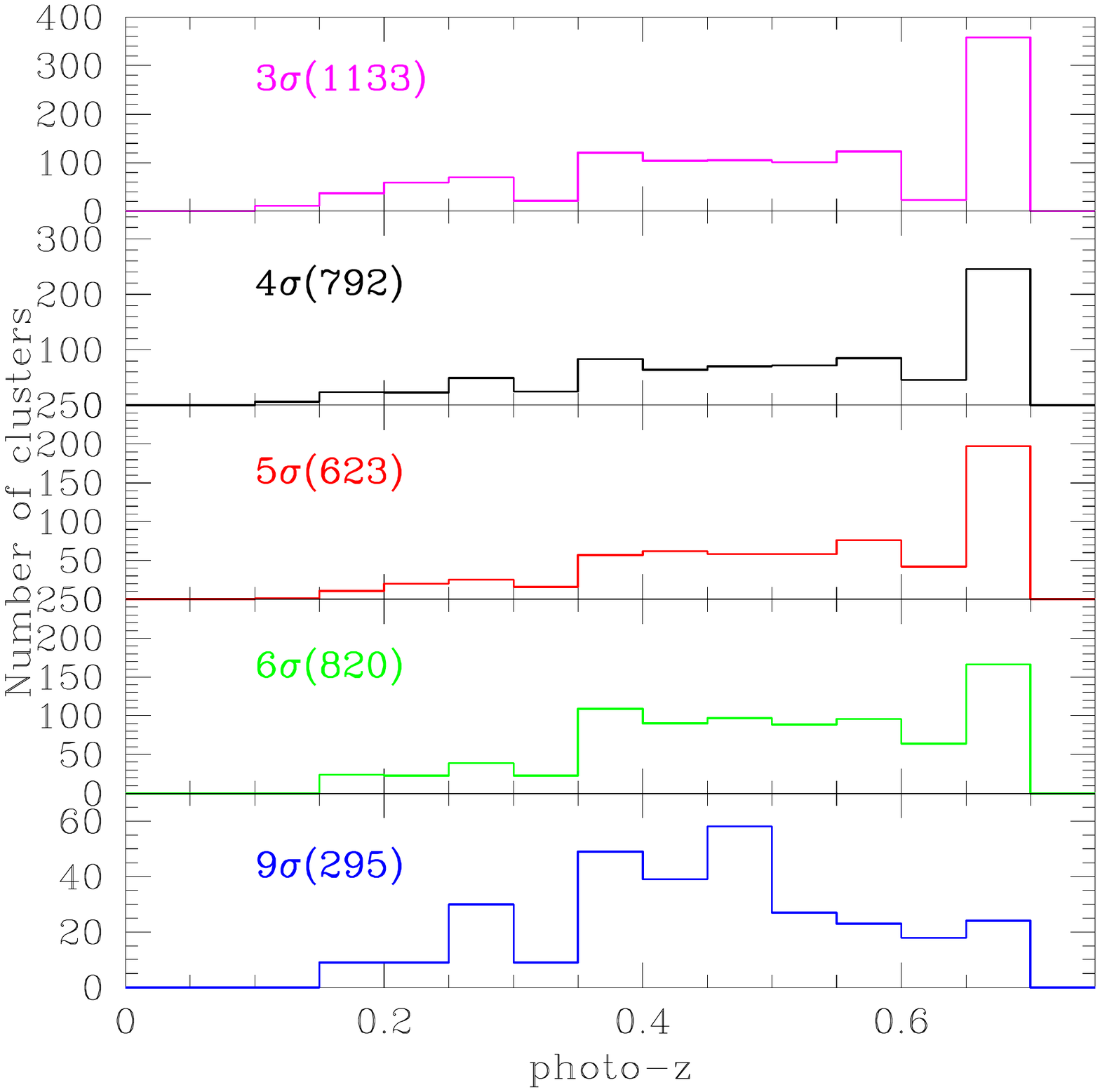}}
  \caption[]{Photometric redshift histograms of the 3663 candidate
  clusters detected in S82 for various detection significance levels
  (indicated in each plot, together with the corresponding number of
  clusters in parentheses). For clarity, the scale of the y-axis is
  not the same for each plot.}
\label{fig:histoz_sig}
\end{figure}

The photo$-z$ histograms for clusters detected at different
significance levels are shown in Fig.~\ref{fig:histoz_sig}.

\subsection{Cluster masses}
\label{sec:clmass}

By applying the same detection method to the Millennium simulation, we
have shown (see A10, Table~2) that there is a rough correspondence
between the cluster detection level and its mass.  We have redone the
same exercise, selecting data from the Millennium simulation and
adapting them to the conditions of the S82 data analysed here, in
terms of photometric redshift precision and photometric catalogue  depth.

\begin{centering}
\begin{table}
  \caption{Percentages of Millennium haloes detected with our method
    as a function of lower mass cutoff.  The columns are: (1)~halo
    mass in units of $10^{12}$~M$_\odot$; (2)--(7)~percentages of
    redetected haloes in the following redshift intervals:
    $z1: z\leq 0.2$, $z2: 0.2<z\leq 0.3$, $z3: 0.3<z\leq
    0.4$, $z4: 0.4<z\leq 0.5$, $z5: 0.5<z\leq 0.6$, and $z6: 0.6< z\leq 0.8$.  }
\begin{tabular}{rrrrrrr}
\hline
\hline
${\rm M_{halo}}$ & $z1$ & $z2$ & $z3$ & $z4$ & $z5$ & $z6$  \\
\hline
250~~  &  -  & 100  &  66  & 33 & 20 & 0 \\
 65~~  &  66 &  68  &  54  & 10 & 5  & 0 \\
 20~~  &  75 &  68  &  59  &  2 & 1 & 0 \\
  7.5  &  76  & 63  &  67  &  2 & 1 & 0 \\
  3.0  &  67  & 64  &  59  &  1 & 1 & 0\\
\hline
\end{tabular}
\label{tab:simu}
\end{table}
\end{centering}

We ran AMACFI on this catalogue, exactly in the same way as for the
S82 galaxy catalogue, and detected 30 structures. The percentages of
detected haloes are given in Table~\ref{tab:simu} for five classes
with masses ranging from $3.0 \times 10^{12}$~M$_\odot$ to $2.5 \times
10^{14}$~M$_\odot$ in six redshift bins: $z\leq 0.2$, $0.2<z\leq 0.3$,
$0.3<z\leq 0.4$, $0.4<z\leq 0.5$, $0.5<z\leq 0.6$, and $0.6< z\leq
0.8$.

We can see that for all the haloes the percentage of detections is
larger than about 60\% up to $z\sim 0.4$.  In the next redshift bin,
this percentage drops to 33\% and 10\% for the two most massive haloes
 and becomes extremely low for the three least massive
haloes.  The corresponding orders of magnitude for the masses are that
clusters detected at 3$\sigma$ and 4$\sigma$ have masses in the
approximate range [$10^{13} - 10^{14}$~M$_\odot$], while clusters
detected at 6$\sigma$ have masses larger than $10^{14}$~M$_\odot$. As
in A10, because the Millennium simulation only covers an
area corresponding to 1~deg$^2$, it includes no cluster corresponding
to a 9$\sigma$ detection in our study, so we cannot estimate the
typical mass of the clusters detected at a 9$\sigma$ level; all we can
say is that these clusters must have masses larger than
$10^{14}$~M$_\odot$.

By varying the detection parameters used in SExtractor, we estimate that
the errors on these halo masses are of the order of 5\%.

\subsection{Cluster spatial density}
\label{sec:cluspadens}

 We found 3663 clusters in a region of about 270~deg$^2$ in the
redshift range 0.15--0.70, which gives a detection rate of about 13.6
clusters per square degree.  Geach et al. (2011) detected 3896 clusters in the same
redshift range, corresponding to about 14.4 clusters per square
degree, a detection rate 1.06 times higher than ours.  This small
difference is most probably due to the fact that they call
``a cluster'' any structure with five galaxies or more. The application of
our cluster detection method to the Millennium simulation shows that
the minimum mass of a $3\sigma$ detected cluster is $\sim
10^{13}$~M$_\odot$, and we  therefore do not detect less massive
structures.

We can also compare the cluster density that we find in S82 with that
found in the four CFHTLS Wide fields. In these fields, we have
detected 4061 candidate clusters at 3$\sigma$ and above, corresponding
to between 21 and 28 clusters per square degree (depending on the field
considered), reaching redshift 1.15 (see D11). The corresponding
cluster densities for $0.15\leq z\leq 0.7$ are between 14.8 and 17.0
clusters per square degree.  The cluster density detected in S82 is
therefore of the same order of magnitude as in the CFHTLS, as seen
from the comparison of Fig.~\ref{fig:histoz_clu} in the present paper
with Fig.~7 (bottom) in D11.

\subsection{Number and magnitude distributions of the cluster galaxies}
\label{sec:maggalinclu}

\begin{figure}[h!]
\centering
  \resizebox{\hsize}{!}{\includegraphics[viewport=10  150 590 710,clip]{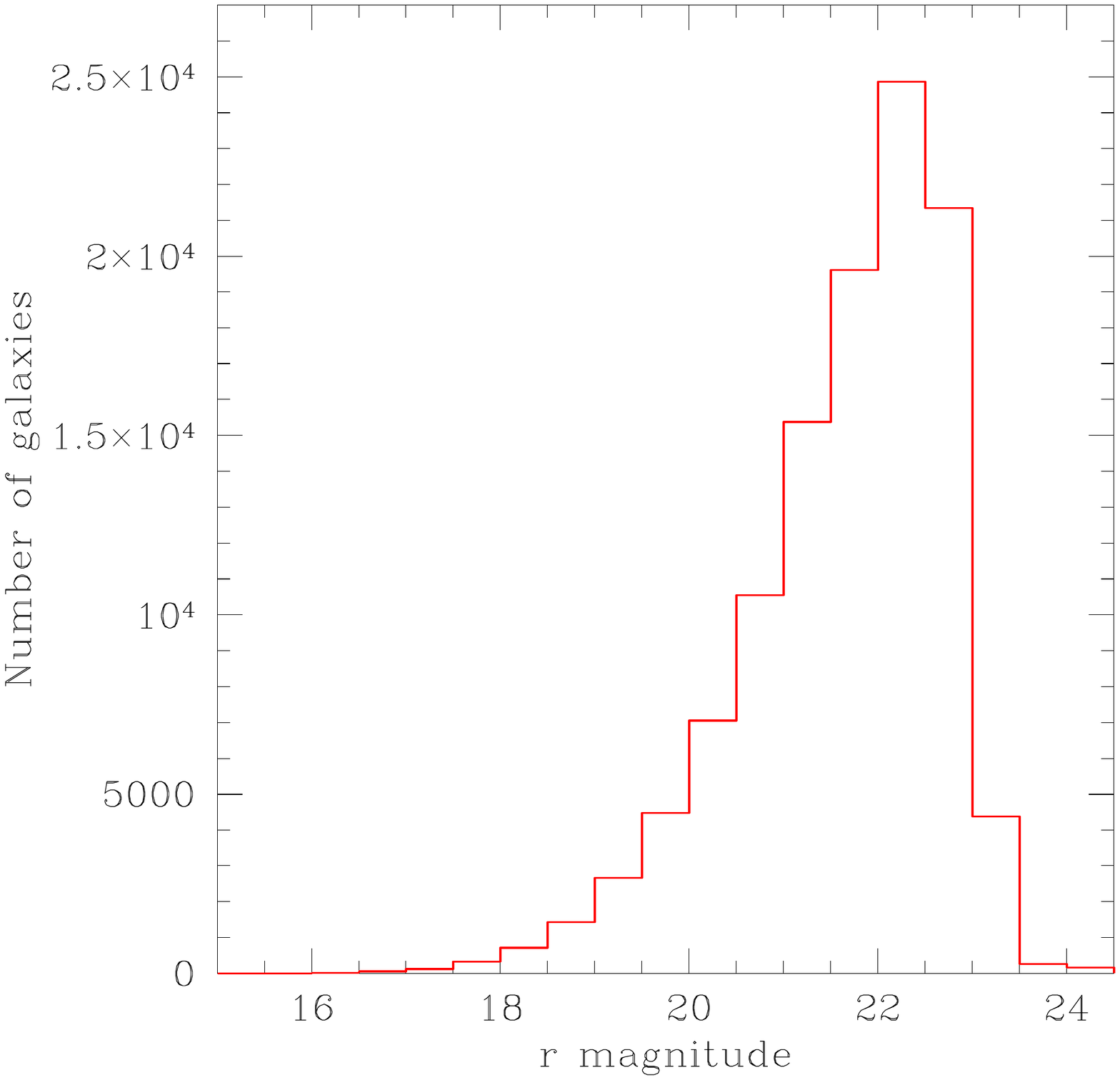}}
  \caption[]{Magnitude histogram of the cluster galaxies in the r band.}
\label{fig:histo_magr_gal}
\end{figure}

We define  cluster galaxies as the galaxies located
within a circle of 1~Mpc radius around each cluster and within $\pm
0.05$ of the redshift of the cluster to which they belong. The
magnitude histogram of the 113,411 cluster galaxies in the r
band is shown in   Fig.~\ref{fig:histo_magr_gal}.

\section{Comparison with optically and X-ray detected clusters in S82}
\label{sec:comparison}

\begin{table}
  \caption{Common systems in our cluster catalogue and in other optically and X-ray 
    selected cluster catalogues within a search radius of 2~Mpc. 
    The percentages of recovered systems
    in the published catalogues are given in parentheses. }
\begin{tabular}{lcrrr}
\hline
\hline
Catalogue   & $z_{phot}$ range & N$_{ClG}$ &  N$_{ClG,match}$ &  N$_{ClG,match}$ \\
         &                  &          & $\Delta z<0.05$ &$\Delta z<0.1$\\
\hline
GMB       & $0.15\leq z_p \leq 0.70$  & 3896 & 472 (12\%) & 838 (22\%) \\
WHL12     & $0.15\leq z_p \leq 0.70$  & 2901 & 538 (19\%) & 838 (29\%) \\
RedMaPPer & $0.15\leq z_p \leq 0.55$  &  665 & 188 (28\%) & 268 (40\%) \\
XCS-DR1   & $0.15\leq z_p \leq 0.70$  &   28 &   5 (18\%) &   7 (25\%) \\
XMM/SDSS  & $0.15\leq z_p \leq 0.68$  &   30 &   5 (17\%) &   6 (20\%) \\
\hline
\end{tabular}
\label{tab:compar}
\end{table}

We have cross-correlated our catalogue of candidate clusters with
several catalogues extracted from optical and/or X-ray data: GMB,
WHL12 (Wen et al. 2012), RedMaPPer (Rykoff et al. 2014), XCS-DR1
(Mehrtens et al. 2012), and XMM/SDSS (Takey et al. 2013, 2014).  The
matching criteria were a linear separation smaller than 2~Mpc and a
redshift difference smaller than 0.05 or 0.1 (see e.g. Hao et
al. 2010).  The numbers of clusters in common are given in
Table~\ref{tab:compar}.

Geach et al. (2011) detected 4098 clusters in the S82 region, but with
a different definition, since they consider that a cluster begins with
five galaxies. The number of recovered clusters from the GMB catalogue
is 22\%, a rather low number.  This can be explained by the fact that
73\% of the GMB clusters have less than ten members, while the
clusters in our sample with the lowest richness have several tens of
galaxies.

The rather low number of recovered clusters from the WHL12 catalogue
(29\%) can be explained in the same way: 30\% of the clusters in WHL12
have 10 members or fewer (in the $r'_{200}$ radius), and 86\% have 20
members or fewer, so WHL12 detect clusters that are mostly less massive
than ours.

In the RedMaPPer catalogue, Ryckoff et al. (2014) give two parameters
($\Lambda$ and S) that can be used to determine the number of cluster
galaxies $N=\Lambda /S$, where $N$ is in the range 20--203.  About
64\% of their clusters have $N<30$ and their cluster masses are
M$_{r200}\geq 10^{14}$ M$_\odot$.  We recover 40\% of their clusters.

Since the detection of our candidate clusters as diffuse X-ray sources
would be an obvious way to confirm that they are real clusters, we
also correlate our detections with the XCS-DR1 catalogue (Mehrtens et
al. 2012). This X-ray catalogue has 41 clusters in the S82 region (as
defined in Section 1), among which 28 are in the same redshift range
as ours.  Our matching percentage is 25\%.

A similar survey to the XCS is the 2XMMi/SDSS galaxy cluster survey
(Takey et al. 2011, 2013, 2014) that provided 35 clusters in the S82
region. Of these, 30 clusters are almost in the same redshift range
[0.15--0.68] as our S82 cluster candidates. 
About 70\% of these clusters  have masses M$_{500} < 10^{14}$~M$_\odot$. With
our cross-matching criteria, we have recovered 20\% of the 2XMMi/SDSS
clusters that are in the S82 region and in the redshift range 
0.15--0.68.

Other observational biases can, however, be present. X-ray serendipitous
surveys such as the XCS and 2XMMi/SDSS make use of existing XMM
observations for which the main targets are most of the time not the
detected clusters. For example, bright stars or large nearby galaxies
can have been targeted, and this would obviously result in a large
masking percentage of the S82 optical data, potentially preventing us
from detecting the X-ray extended structure as a galaxy
concentration. In addition, these serendipitous surveys of clusters
avoided the clusters (usually the massive ones) that were targets of
pointed XMM-Newton observations. All these observational biases reduce
the recovered fraction of the X-ray selected clusters. It is worth 
performing a detailed comparison of X-ray clusters and our S82 clusters similar to
the X-CLASS-redMaPPer galaxy cluster comparison (Sadibekova et
al. 2014). We therefore plan in the near future to make this detailed
comparison.

\section{Properties of stacked clusters}
\label{sec:properties}

In this section we will limit our analysis to the 1738 clusters
detected at 5$\sigma$ and above, and to galaxies within a radius of
1~Mpc of a cluster and with a photo$-z$ within $\pm 0.05$ of that of
the corresponding cluster for two  reasons: first, to avoid having
too much contamination by galaxies that do not belong to the
clusters, and second to derive galaxy luminosity functions (GLFs) in
redshift bins of width 0.1 that do not overlap.

\subsection{Colour-magnitude diagrams}
\label{sec:colmag}

\begin{figure*}[t!]
\centering
  \resizebox{5cm}{!}{\includegraphics[viewport=15  150 590 710,clip]{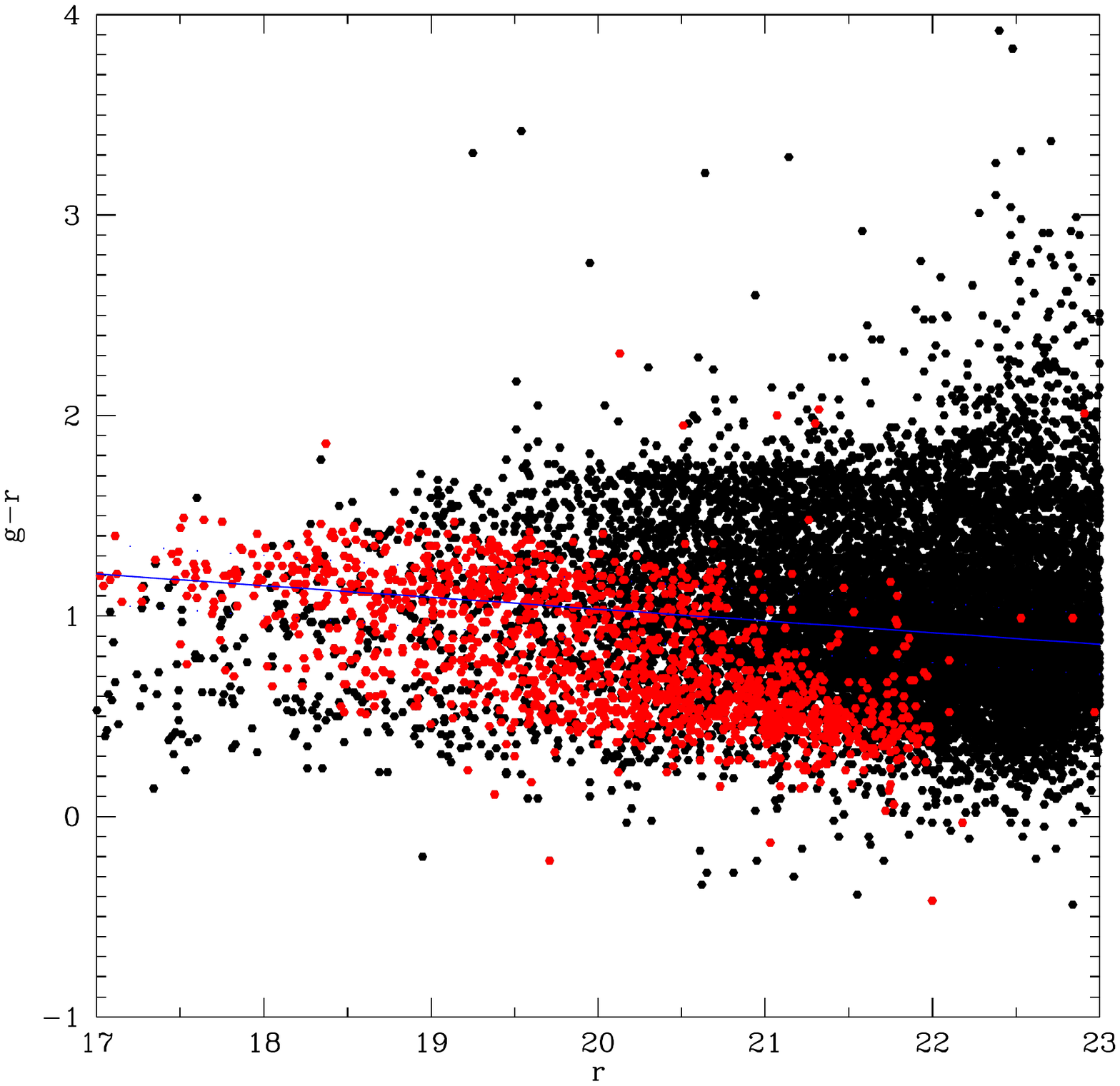}}
  \resizebox{5cm}{!}{\includegraphics[viewport=15  150 590 710,clip]{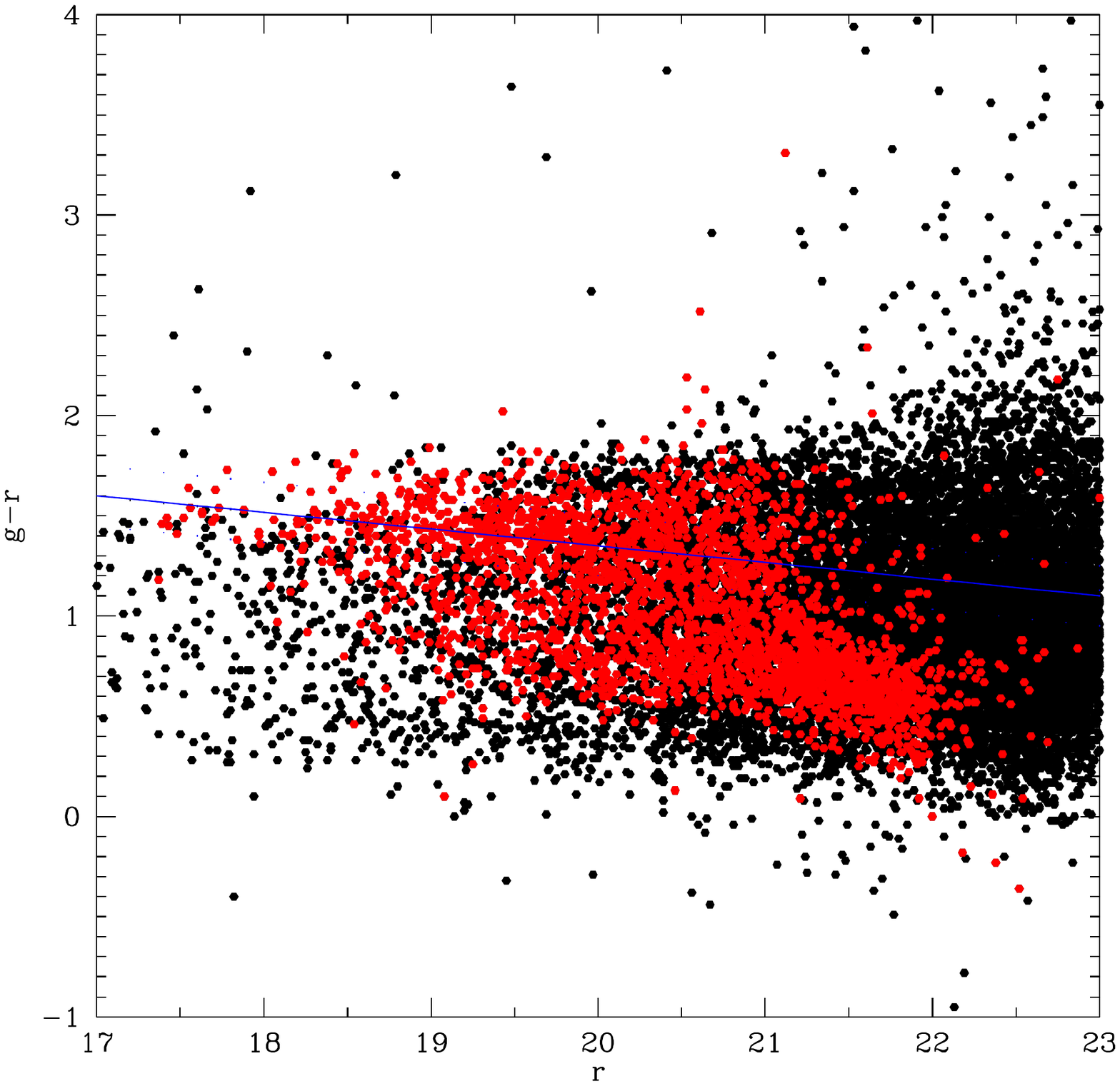}}
  \resizebox{5cm}{!}{\includegraphics[viewport=15  150 590 710,clip]{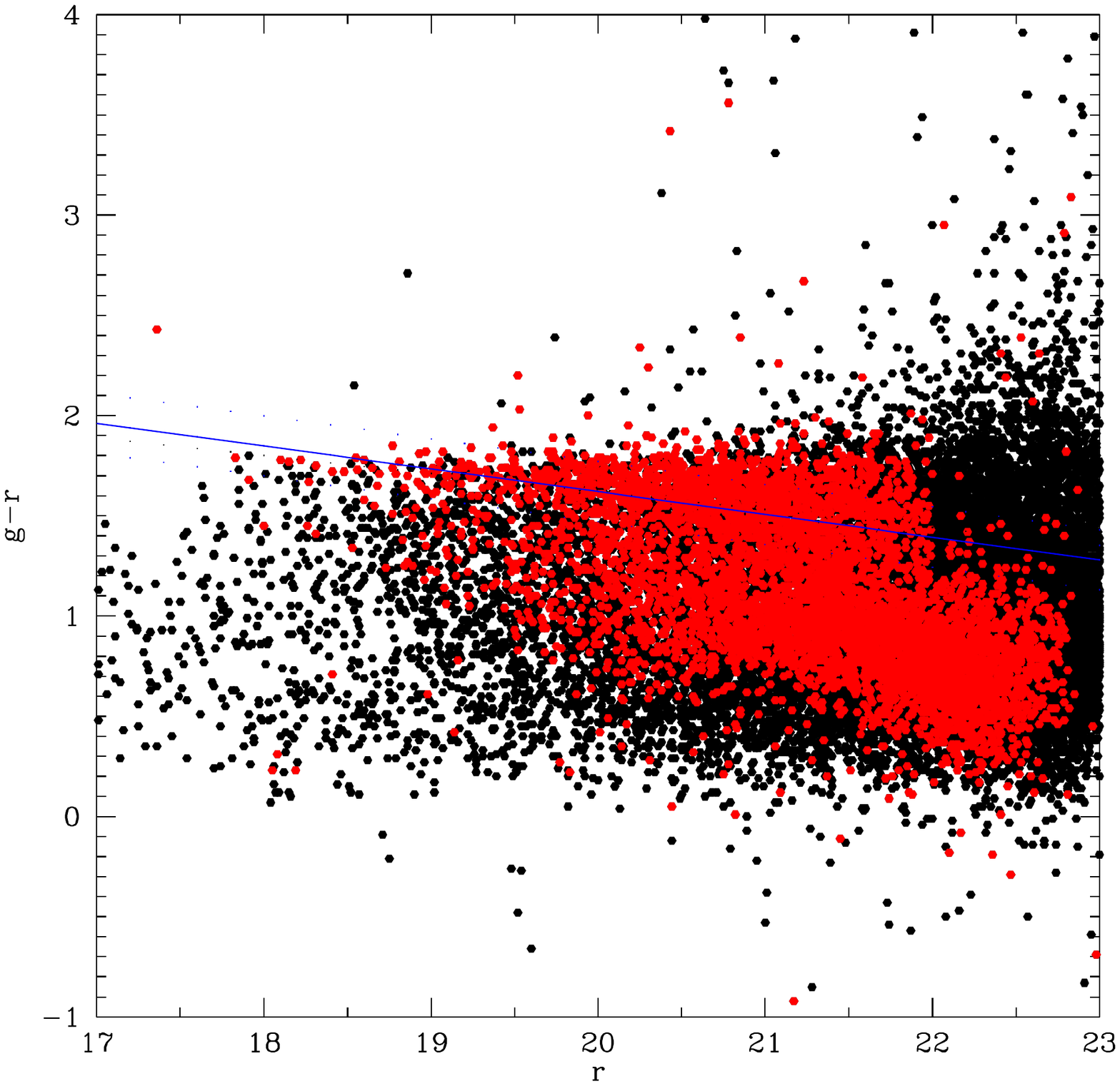}}
  \resizebox{5cm}{!}{\includegraphics[viewport=15  150 590 710,clip]{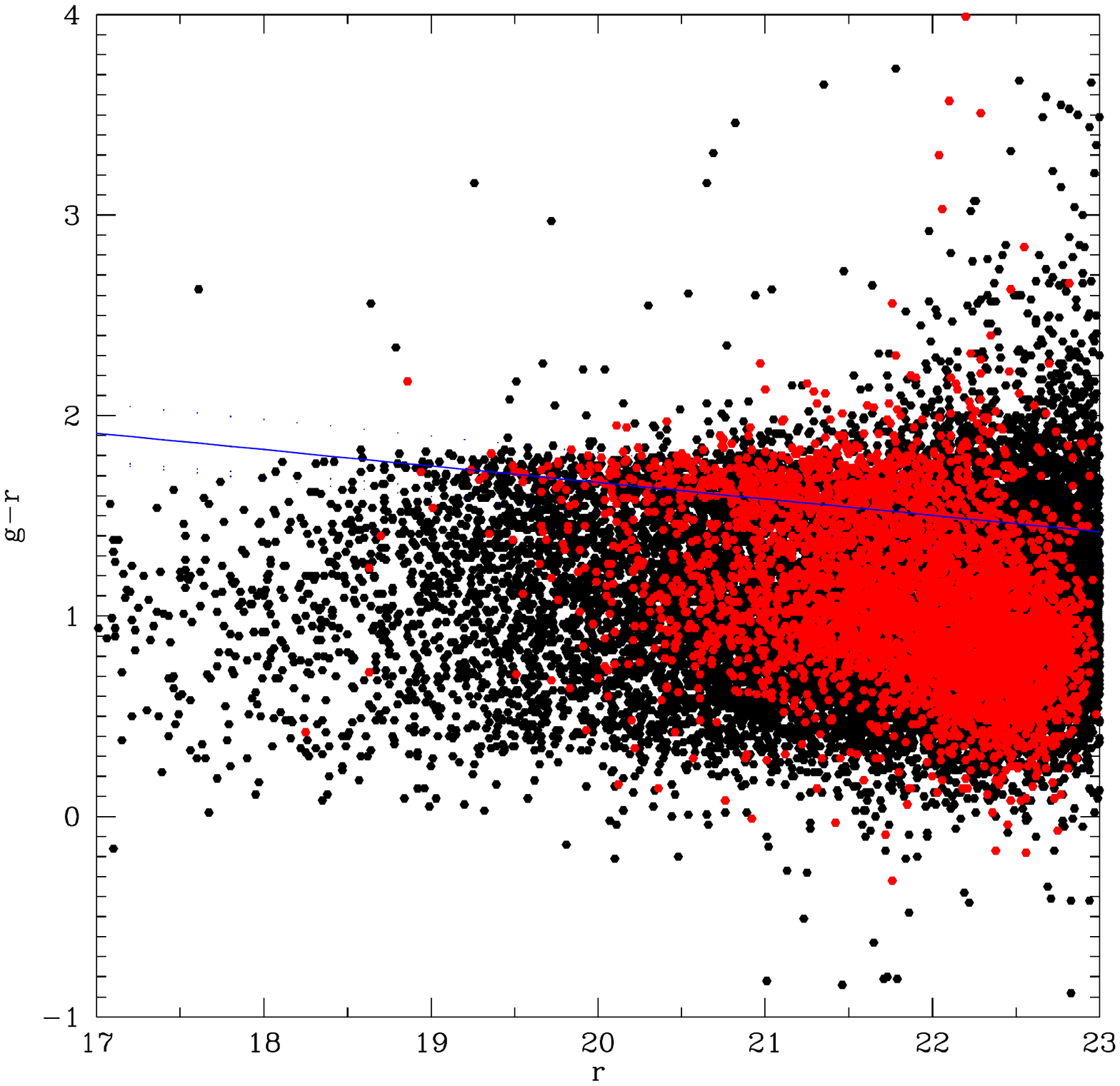}}
  \resizebox{5cm}{!}{\includegraphics[viewport=15  150 590 710,clip]{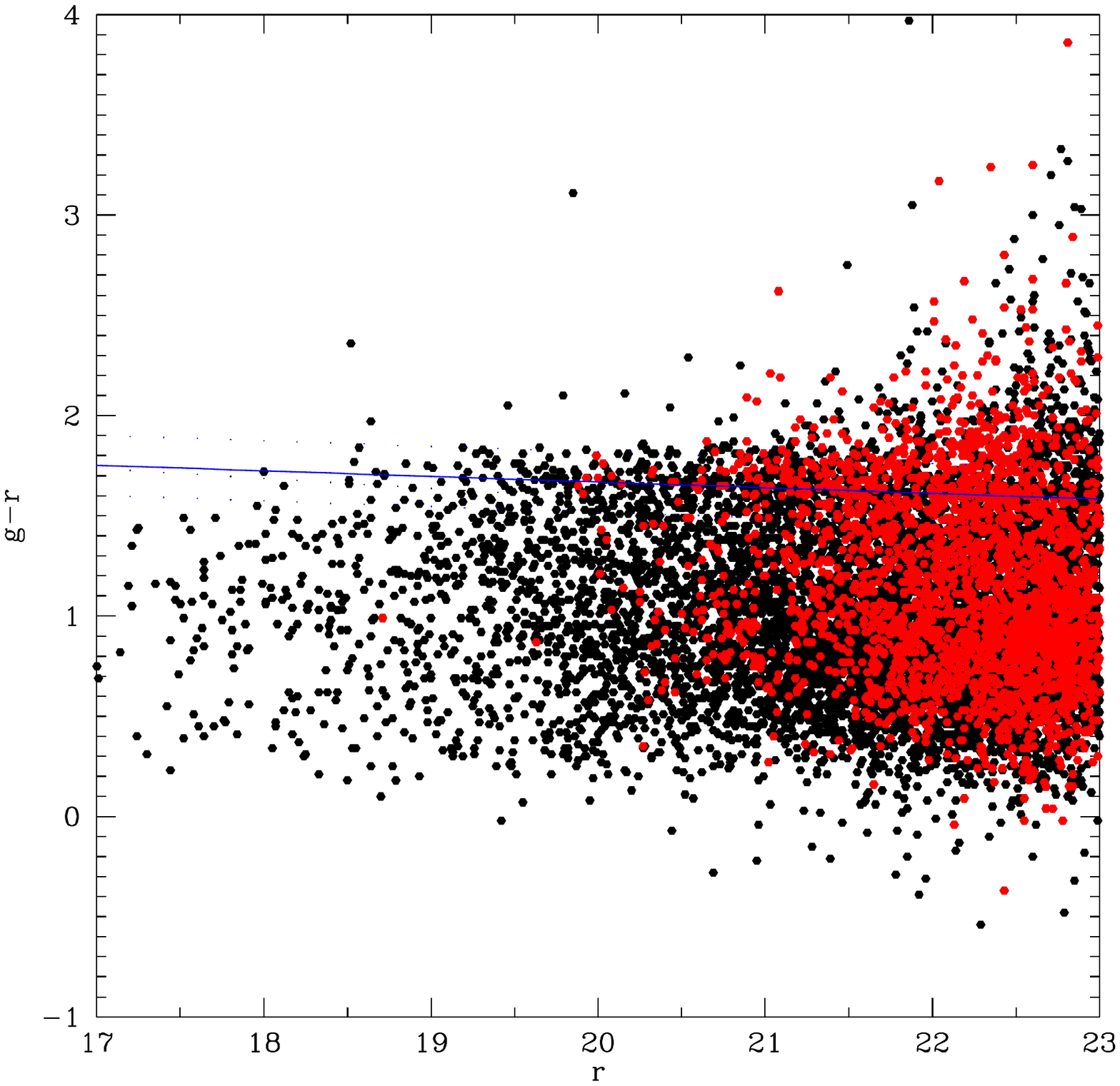}}
\caption[]{Colour-magnitude diagrams ($(g-r)$ versus $r'$) stacked in
  the five redshift bins (from left to right: $0.15<z<0.25$,
  $0.25<z<0.35$, $0.35<z<0.45$, $0.45<z<0.55$, $0.55<z<0.65$) before
  background subtraction (see text). The black points show all the
  galaxies within a radius of 1~Mpc of a cluster, and the red points
  correspond to the galaxies with a photometric redshift within $\pm
  0.05$ of that of the cluster to which they belong. 
  The solid blue line shows the best fit.}
\label{fig:colmag}
\end{figure*}

\begin{figure*}[t!]
\centering
  \resizebox{5cm}{!}{\includegraphics[viewport=15  150 590 710,clip]{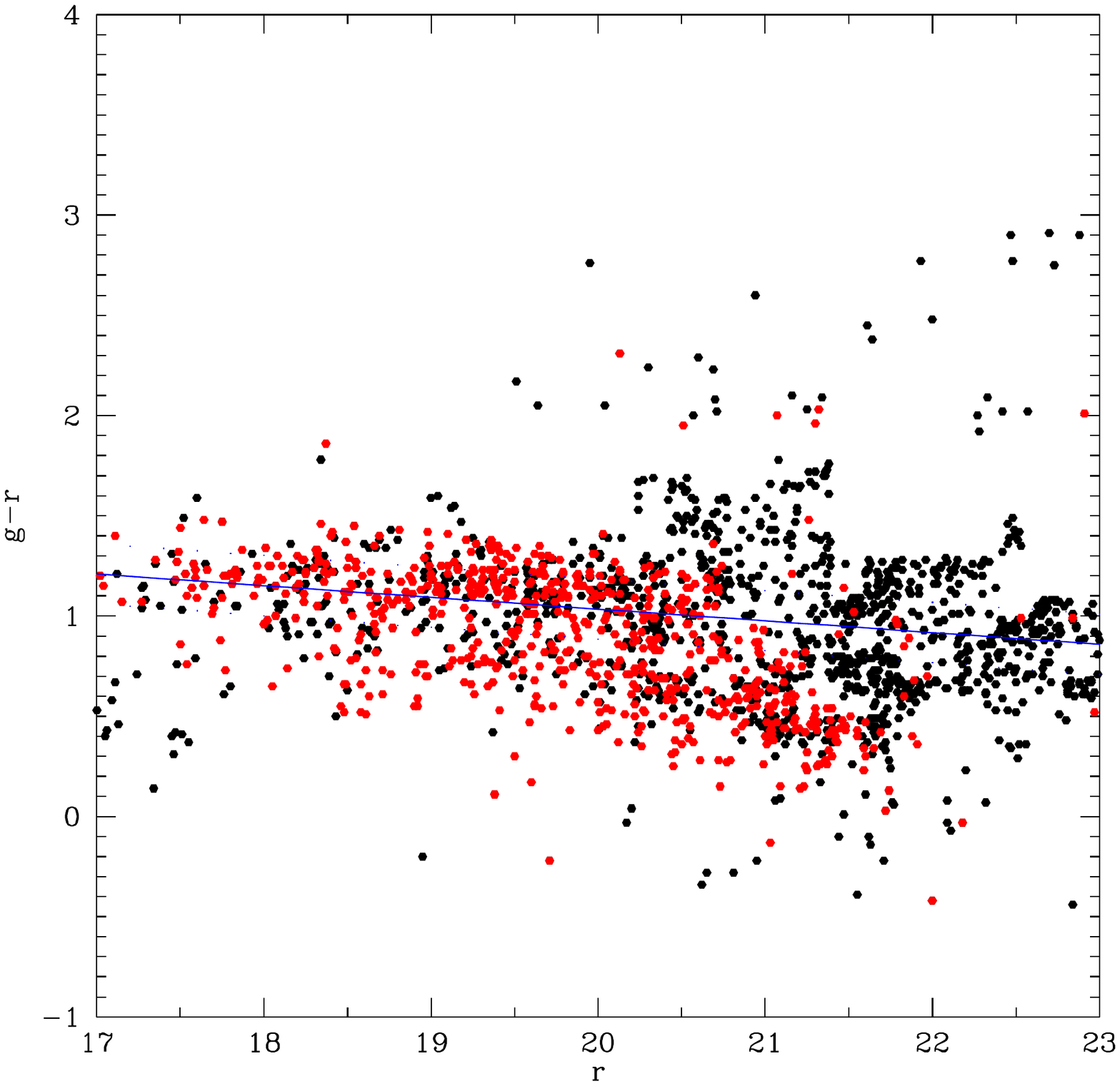}}
  \resizebox{5cm}{!}{\includegraphics[viewport=15  150 590 710,clip]{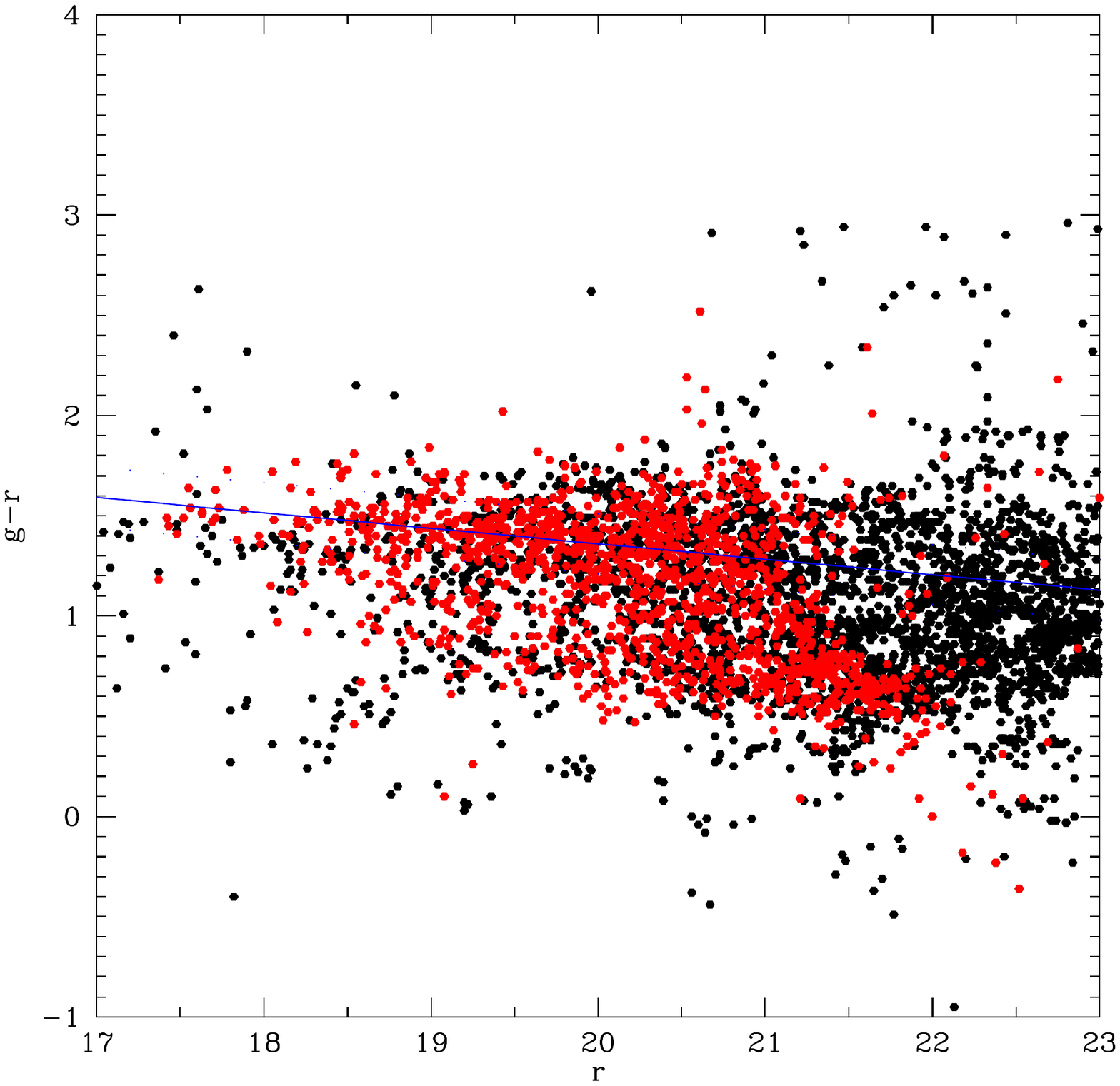}}
  \resizebox{5cm}{!}{\includegraphics[viewport=15  150 590 710,clip]{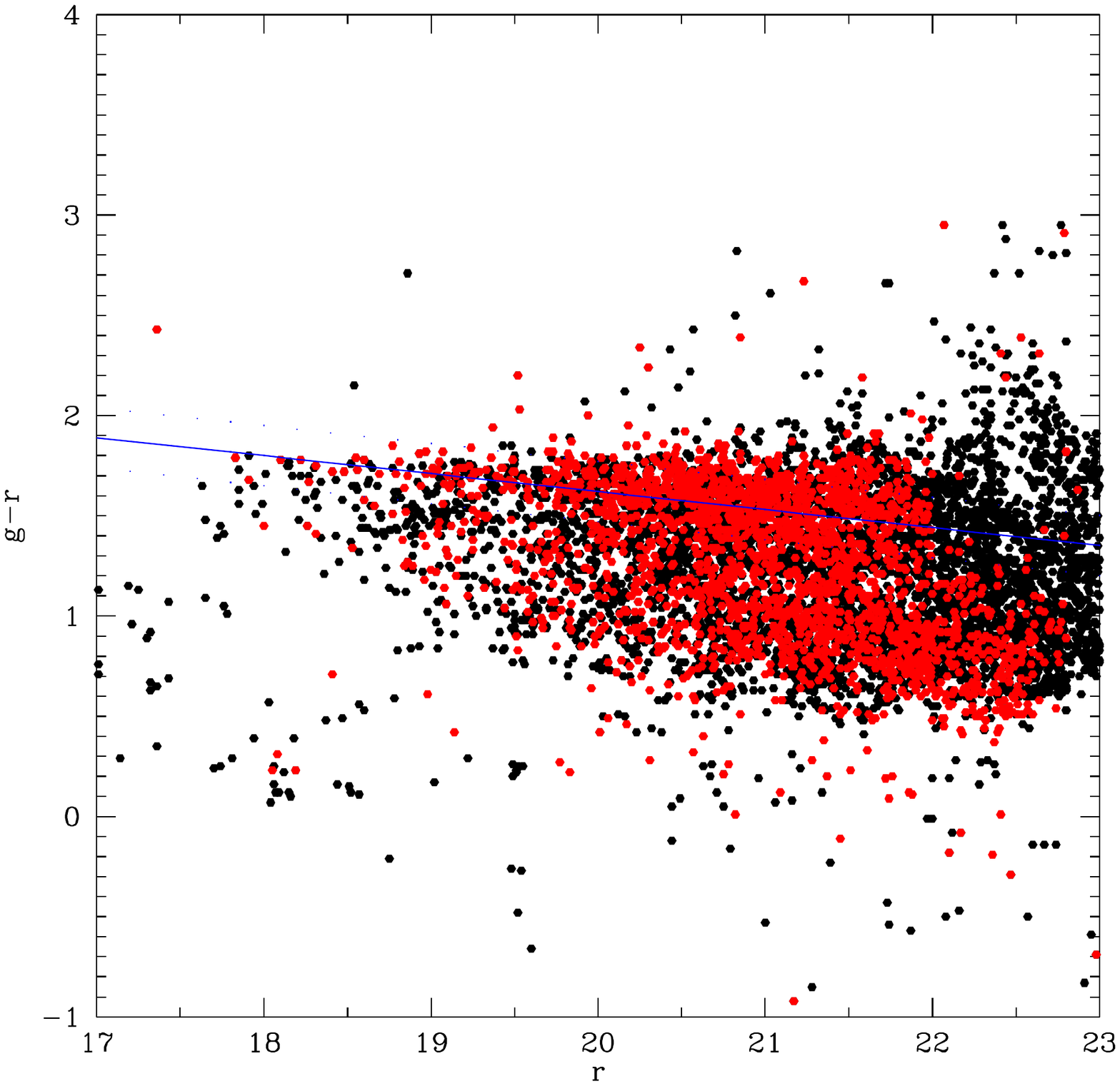}}
  \resizebox{5cm}{!}{\includegraphics[viewport=15  150 590 710,clip]{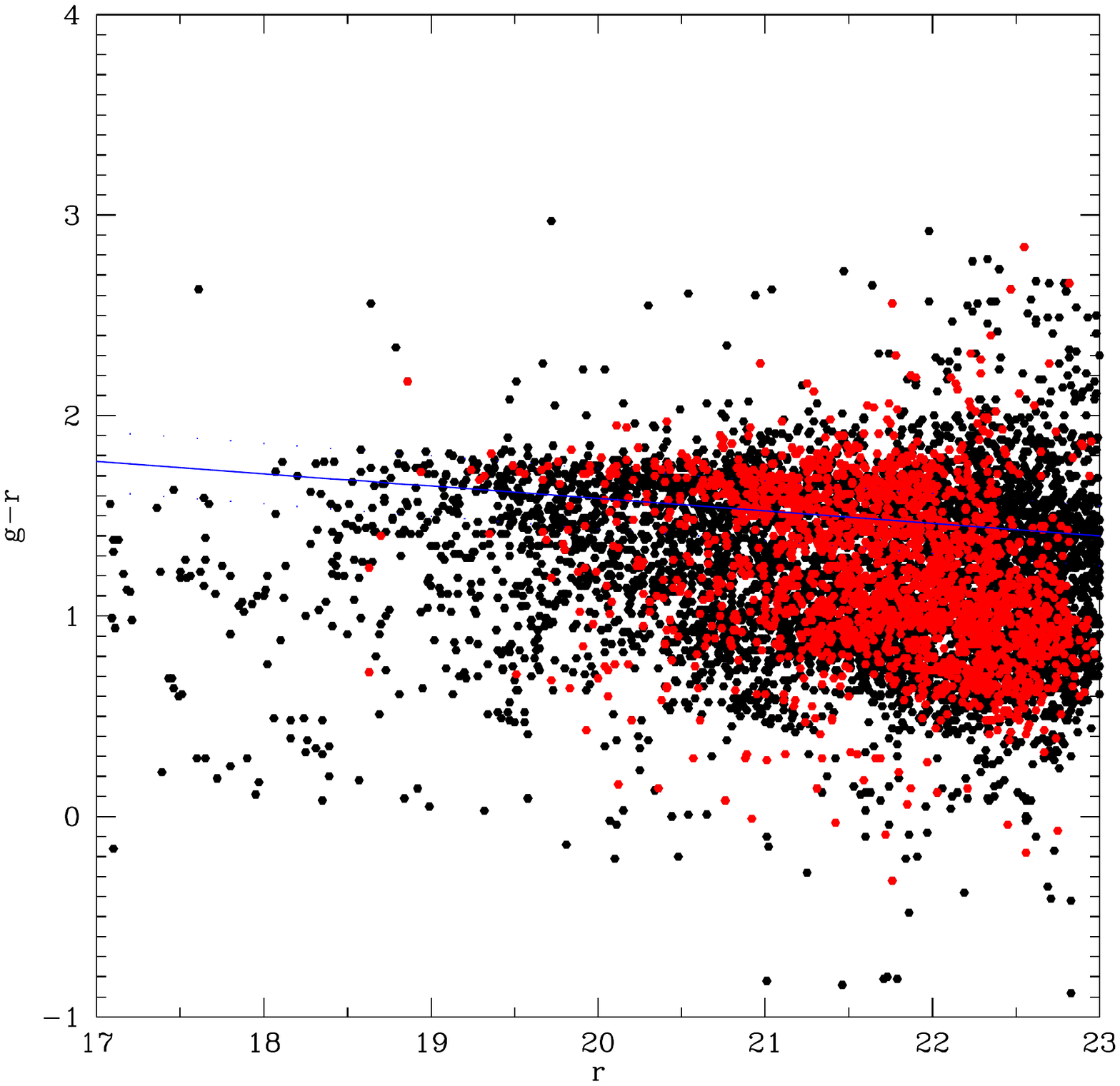}}
  \resizebox{5cm}{!}{\includegraphics[viewport=15  150 590 710,clip]{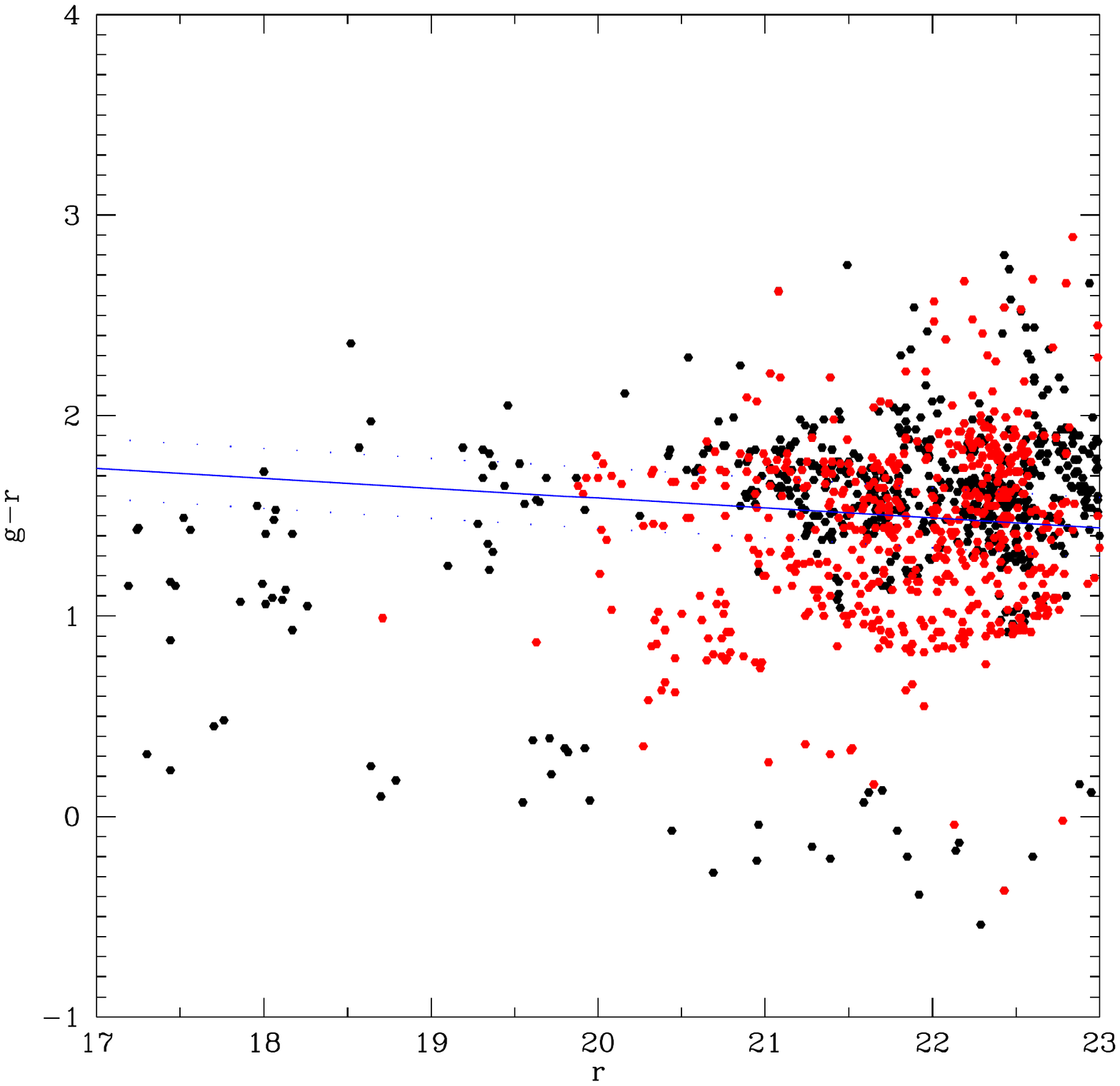}}
\caption[]{Same as Fig.~\ref{fig:colmag} after background subtraction. }
\label{fig:colmagbgcor}
\end{figure*}

We first derive colour-magnitude diagrams by stacking galaxies in
photometric redshift bins. The red sequence is apparent in all
colour-magnitude diagrams, but the $(g-r)$ versus $r'$
colour-magnitude diagram is the one that shows the smallest dispersion,
and so we  use it  to select cluster galaxies and build
GLFs. We show these diagrams in five redshift bins in
Figs.~\ref{fig:colmag} and \ref{fig:colmagbgcor}, respectively, before
and after background correction (see Section~\ref{sec:backsub} for
explanations on the method used to subtract the background contribution).

The fact that we detect a red sequence shows that we  have
selected galaxies with similar star formation histories that belong to
well-assembled structures, and therefore that our candidate clusters
are mostly old galaxy structures.
As seen in these figures, the red sequence defined by the cluster
galaxies is in  good agreement with the predictions of the Bruzual
\& Charlot (2003) model.

\subsection{Galaxy luminosity functions}
\label{sec:GLF}

\begin{figure*}[t!]
\centering
  \resizebox{\hsize}{!}{\includegraphics[viewport=30  460 590 800,clip]{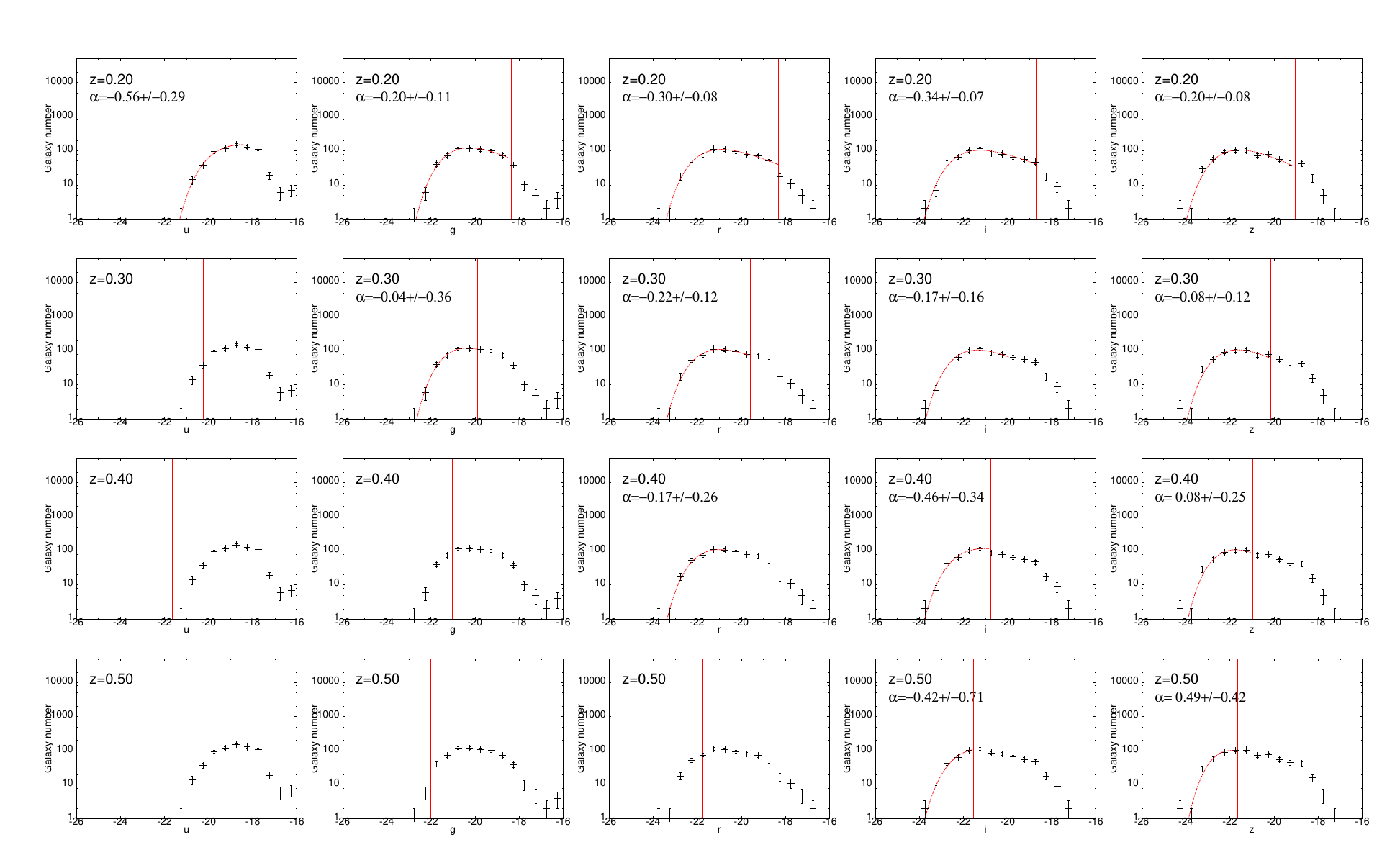}}
  \caption[]{GLFs in the u, g, r, i and z bands (from left to right).
  Black crosses are the stacked GLFs with no surface
  normalization. The red curves are the best Schechter fits
  corresponding to the black crosses. The vertical red lines indicate
  the 90\% completeness limits. Only galaxies brighter than the 90\%
  completeness limit are taken into account for the fits. The plots
  for which no Schechter parameters are given in
  Table~\ref{tab:schechparam} are only shown for completeness, but
  will not be taken into account in the discussion.  }
\label{fig:GLFs}
\end{figure*}

\subsubsection{Computing absolute magnitudes and 90\% completeness limits}

In order to be able to stack the cluster GLFs it was necessary to
compute absolute magnitudes for all the galaxies. For this, we applied
the Le~Phare software as described in Section~\ref{sec:cats}, with
photo$-z$s fixed to the SDSS values.

Since the GLF parameters can strongly vary with the magnitude interval
in which they are computed (as discussed e.g. by Martinet et al. 2015
and references therein), it is necessary to estimate the absolute
magnitudes for which the completeness is better than 90\%.  For this,
our starting point is the 90\% completeness limits given in
Section~2.1: $u'\sim 23.0$, $g'\sim 22.8$, $r'\sim 22.1$, $i'\sim 21.5$,
and $z'\sim 21.2$.  We computed the corresponding 90\% completeness
limits in absolute magnitudes with two independent methods.

First, we translated these apparent magnitude completeness limits to
absolute magnitude completeness limits by applying in each redshift
bin the k-correction and distance modulus.  Le~Phare uses galaxy SED
model libraries to estimate the theoretical k-corrections that depend
on galaxy types and redshifts. For early-type galaxies, we measure the
mean and the dispersion of the k-correction over galaxy templates in a
redshift range of $\pm$0.05 around the cluster redshift.
We  set our corrective factors to the mean values plus 2$\sigma$
to be representative of 95\% of our galaxy population.  This step is
illustrated in Eq.~\ref{eq:kcorr} where $C_X$ and $C_x$ are the
completeness limits in absolute and apparent magnitude in the x band,
$DM(z)$ is the distance modulus, and $k_x(z)$ the k-correction in the x band
at redshift z:

\begin{equation}
\label{eq:kcorr}
C_X = C_{x} - DM(z) - (<k_x(z)> + 2\sigma_{k_x(z)})
.\end{equation}
\noindent
With this method we obtained the 90\% completeness limits in absolute
magnitude for each filter and each redshift bin. These values are
given in the last column of Table~\ref{tab:schechparam}.

As a check, we also performed simulations for 112 clusters. For this,
we first selected in each cluster the galaxies with a photo$-z$ within
$\pm 0.05$ of that of the cluster (i.e. galaxies for which the
distance modulus and k-corrections are known) and with no nearby
neighbour (i.e. no galaxy within 3 times their size) to avoid crowding
effects. We extracted the image of each of these galaxies, subtracted
the background computed around each galaxy at a distance larger than 3
times the galaxy size (as measured by SExtractor), and added this
background subtracted image 100 times at uniformly distributed random
locations within a square of $2000\times 2000$~pixels$^2$ centred on
the cluster centre. We then redetected the galaxy on the image with
SExtractor and noted how many times it was redetected. This allowed us
to estimate the number of times we could redetect a galaxy with
the absolute magnitude of the considered galaxy. By applying this
treatment to all the cluster galaxies, we thus reconstructed a
completeness curve as a function of absolute magnitude.  Since such a
computation is only valid for a small magnitude range, we repeated it
for galaxies 10 times brighter (with magnitudes smaller by 2.5) and 10
times fainter (with magnitudes larger by 2.5), and obtained curves
such as those shown in Fig.~\ref{fig:simucompl} for the $i'$ band.  This
method has obvious limitations, but it gives 90\% completeness limits
very close to those estimated with our first method, in most cases
within one 0.5 magnitude bin, thus giving us confidence in our
completeness level estimates. Hereafter, we will limit our GLF fits to
the 90\% completeness absolute magnitude limits derived with the first 
method.

\subsubsection{Background subtraction}
\label{sec:backsub}

As stated above, we extracted a catalogue containing all the galaxies
located within a 1~Mpc radius around each cluster and with a
photo$-z$ within $\pm 0.05$ of that of the corresponding cluster.
The composite ($g'-r'$) versus $r'$ colour magnitude diagrams have been
corrected for contamination from background/foreground galaxies in a
statistical way. For this purpose, the field colour-magnitude diagram
has been estimated from the whole S82 distribution, excluding galaxies
in a given physical radius (in our case 1~h$^{-1}$~Mpc) around the
position of detected clusters.  The statistical correction has been
performed following the method described in Pimbblet et
al. (2002). Counts in the ``cluster + field'' and ``field''
populations are estimated in a grid in the colour-magnitude diagram,
and the probability of a galaxy in a colour-magnitude bin of being a
field galaxy is derived and used to statistically subtract the field
population. This method has been applied to the composite clusters
stacked in photo$-z$ bins. In the case of subsamples of the stacks
where galaxies are selected in a photometric redshift window around
the cluster mean redshift, a grid in the colour-magnitude-photometric
redshift space is used. More details will be provided in Maurogordato
et al. (in preparation).

After this statistical background subtraction was applied, for each
redshift bin we extracted the galaxies located in the red sequence of
the ($g'-r'$) versus $r'$ colour magnitude diagrams and thus obtained
the GLFs that we fit with a Schechter
function:
\begin{equation}
\label{eq:schech}
 N(M)  = 0.4\ log_{10}\phi^*[10^{0.4(M^*-M)}]\ ^{\alpha+1}exp\ (-10^{0.4(M^*-M)})
.\end{equation}

\subsubsection{Results}

\begin{center}
\begin{table}
\caption{Parameters of the best fit Schechter functions for the
galaxy luminosity functions in the five bands and in the five redshift bins.  }
\label{tab:schechparam}
\begin{tabular}{lcccc}
\hline\hline
redshift/ & $\alpha$ & M* &$\phi$*&90\% \\
filter    &          &    &       & completeness \\
\hline
z=0.20 & & & & \\
$u'$ & $-0.56\pm 0.29$ & $-19.2\pm  0.2$ &   368$\pm$    50 & $-18.4$ \\
$g'$ & $-0.20\pm 0.11$ & $-20.5\pm  0.1$ &   352$\pm$    23 & $-18.4$ \\
$r'$ & $-0.30\pm 0.08$ & $-21.3\pm  0.1$ &   306$\pm$    21 & $-18.3$ \\
$i'$ & $-0.34\pm 0.07$ & $-21.7\pm  0.1$ &   285$\pm$    22 & $-18.7$ \\
$z'$ & $-0.20\pm 0.08$ & $-21.8\pm  0.1$ &   304$\pm$    19 & $-19.0$ \\
\hline
z=0.30 & & & & \\
$u'$ & $ -$ & $-$ & $-$ & $-20.2$ \\
$g'$ & $-0.04\pm$ 0.36 & $-20.4\pm$  0.2 &   353$\pm$    24 & $-19.9$ \\
$r'$ & $-0.22\pm$ 0.12 & $-21.2\pm$  0.1 &   314$\pm$    22 & $-19.6$ \\
$i'$ & $-0.17\pm$ 0.16 & $-21.6\pm$  0.1 &   307$\pm$    23 & $-19.8$ \\
$z'$ & $-0.08\pm$ 0.12 & $-21.7\pm$  0.1 &   314$\pm$    18 & $-20.1$ \\
\hline
z=0.40 & & & & \\
$u'$ &$-$             & $-$             & $-$              & $-21.6$ \\
$g'$ &$-$             & $-$             & $-$              & $-21.0$ \\
$r'$ &$-0.17\pm$ 0.26 & $-21.2\pm$  0.2 &   315$\pm$    21 & $-20.7$ \\
$i'$ &$-0.46\pm$ 0.34 & $-21.7\pm$  0.2 &   302$\pm$    41 & $-20.8$ \\
$z'$ &$ 0.08\pm$ 0.25 & $-21.7\pm$  0.1 &   315$\pm$    19 & $-21.0$ \\
\hline
z=0.50 & & & & \\
$u'$ &$-$             & $-$             & $-$              & $-22.9$ \\
$g'$ &$-$             & $-$             & $-$              & $-22.0$ \\
$r'$ & $-$ & $-$ & $-$ & $-21.8$ \\
$i'$ &$-0.42\pm$ 0.71 & $-21.7 \pm$  0.4 &   303$\pm$   39 & $-21.5$ \\
$z'$ & 0.49$\pm$ 0.42 & $-21.5\pm$  0.2  &   267$\pm$    54 & $-21.6$ \\
\hline
\end{tabular}
\label{tab:shechter}
\end{table}
\end{center}

We stacked the 1738 clusters detected at 5$\sigma$ and above, limiting
our analysis to the galaxies within a radius of 1~Mpc of a cluster and
with a photo$-z$ within $\pm 0.05$ of that of the corresponding
cluster and subtracting the background as explained above. This
allowed us to obtain stacked GLFs in the same five redshift bins as
for the colour-magnitude diagrams.
   
The GLFs in the $u', g', r', i'$, and $z'$ bands are shown in
Fig.~\ref{fig:GLFs} and the parameters of the best fit Schechter
functions are given in Table~\ref{tab:shechter}.  No values were given
in Table~\ref{tab:shechter} when the fits did not converge.  This
happened mostly when the number of points brighter than the 90\%
completeness limit became too small for a three-parameter fit.  In
some cases, the fits converged, but gave values with large error
bars. We chose to show these values in Table~\ref{tab:shechter} to
keep the information as complete as possible, but  they
should be considered with caution. 

In Fig.~\ref{fig:GLFs} it  can be seen that in some cases
there is an excess of very bright galaxies over the Schechter
function, mostly in the $z'$ band. This feature
is rather common, particularly in merging clusters (see e.g. Durret et
al. 2011a and references therein).   We checked the
possibility that this excess could be due to bright stars
misclassified as galaxies in one cluster.  For
this we detected all the objects with SExtractor in the $i'$ band image
and plotted the maximum surface brightness as a function of magnitude
($\mu _{\rm max,i}-i$ diagram). The bright objects from our initial
galaxy catalogue that could account for the excess of bright galaxies
in the GLF are all very clearly located in the galaxy zone in the $\mu
_{\rm max,i}-i$ diagram, so it seems likely that the excess of very
bright galaxies detected in some cases is real, and not due to bright
stars misclassified as galaxies.

If we now consider the faint end slope of the GLF, we can see that
$\alpha$ is above $-1$, traducing a decrease in the faint galaxy
population, and this drop becomes more significant with increasing
redshift, at least in the bands where the fit converges in the highest
redshift bins. As expected from the relative shallowness of the images
in the $u'$ band, the GLFs can only be computed in the first redshift
bin.

At low redshifts there are fewer faint galaxies than expected
($\alpha$ is notably larger than the expected value of $\sim -1$),
probably in part due to background contamination.  The $\alpha$
parameter of early-type field GLFs is about $-0.16$ in U and $-0.31$
in the V, R and I bands in the redshift range $0.4<z<0.9$, and is
found to depend only weakly on redshift (Zucca et al. 2006).

\section{Morphological properties of cluster galaxies}
\label{sec:propgal}

\subsection{Early- to late-type galaxy fraction}

Based on the catalogue of clusters that we have detected in a
homogeneous way, we now analyse statistically the morphological
properties of the galaxies belonging to these clusters (or at least
having a high probability of being in these clusters, since this study
is based on photometric redshifts).  With the large number of cluster
galaxies available, this allows us to estimate the variations of the
late- to early-type number ratio as a function of redshift and of
detection level. Because the positions of the cluster centres are not
well defined, we will not attempt to search for variations of the
elliptical-to-spiral number ratio as a function of clustercentric
radius.  We consider here the cluster galaxies, with the definition
given in Section~\ref{sec:properties}.

To estimate the morphological properties of the galaxies, we extracted
images around each cluster, covering an area of $1\times 1$~Mpc$^2$ at
the cluster redshift, with a pixel scale of 0.396~arcsec/pixel, in the
$r'$ band.  We applied a tool developed in SExtractor that calculates
the respective fluxes in the bulge (spheroid) and disk for each
galaxy. This new experimental {\sc SExtractor} feature fits to each
galaxy a two-dimensional model comprised of a de Vaucouleurs spheroid
(the bulge) and an exponential disk. Briefly, the fitting process is
very similar to that of the GalFit package (Peng et al. 2002) and is
based on a modified Levenberg-Marquardt minimization algorithm. The
model is convolved with a supersampled model of the local point spread
function (PSF), and downsampled to the final image resolution.  The
PSF model used in the fit was derived with the {\sc PSFEx} software
(Bertin 2011) from a selection of point source images.  The PSF
variations were fit using a six--degree polynomial of $x$ and $y$
image coordinates. The model fitting was carried out in the $r'$ band.

We used this tool to look for differences in galaxy morphologies as
a function of redshift and of significance level (which is related to
cluster mass) by computing  the flux in the disk
$f_{disk}$ and that in the spheroid $f_{spheroid}$ for each galaxy. We classified
  a galaxy as early-type if $f_{spheroid} / (f_{disk}+f_{spheroid})
  \geq 0.35$ and as late-type if $f_{spheroid} /
  (f_{disk}+f_{spheroid}) < 0.35$, as in Simard et
  al. (2009). SExtractor also computes the $1\sigma$ uncertainties on
these fluxes and on the $f_{spheroid} / (f_{disk}+f_{spheroid})$ flux
ratio. We must note that the
distribution of the estimated uncertainties can be highly asymmetric
and that the limiting value of 0.35 for the $f_{spheroid} /
(f_{disk}+f_{spheroid})$ ratio to distinguish early and late types is
somewhat arbitrary (see e.g. Simard et al. 2009 and references
therein).

\begin{figure}
\centering
  \resizebox{\hsize}{!}{\includegraphics[viewport=0  15 550 325,clip]{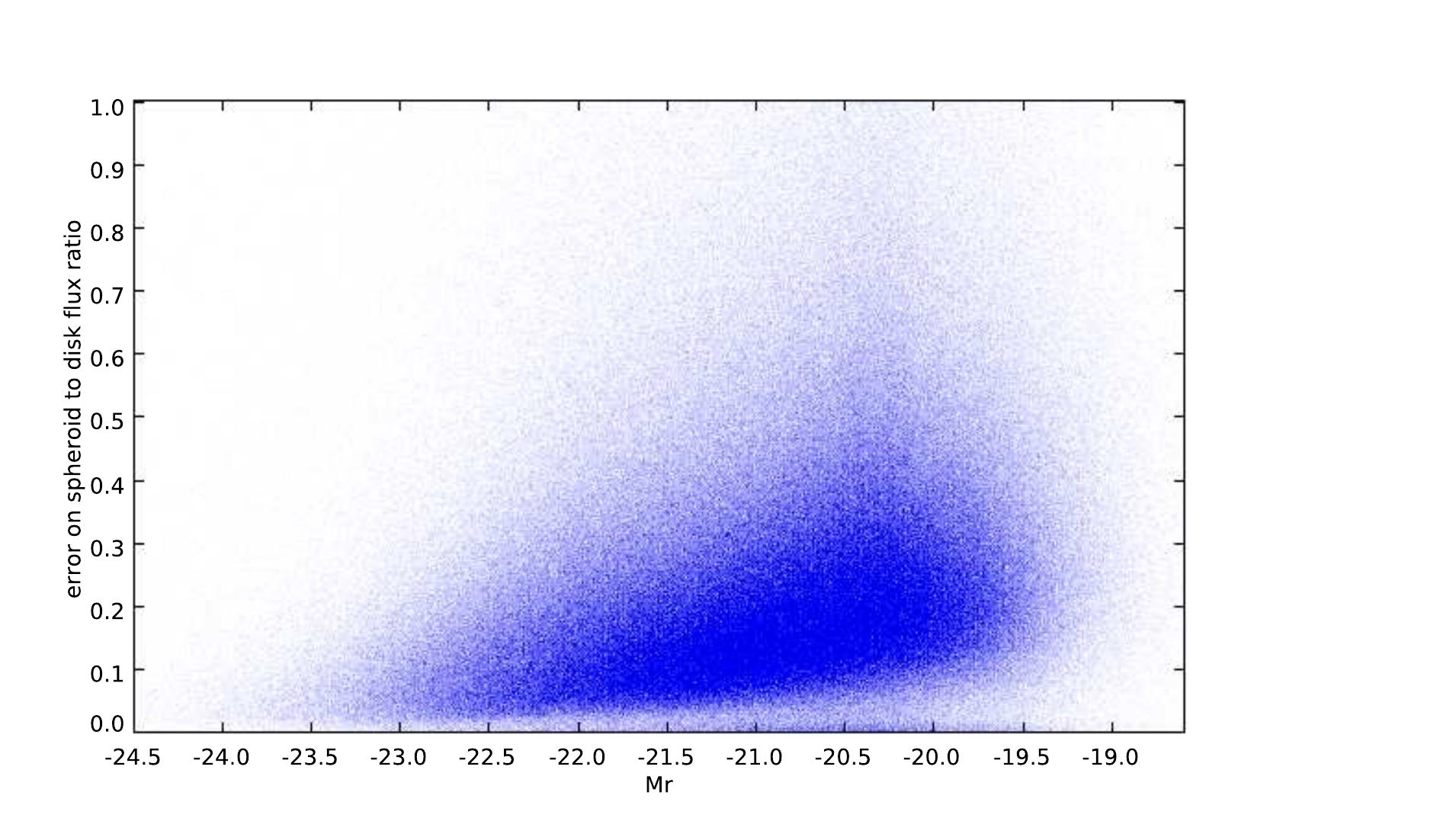}}
  \caption[]{Relative error on the spheroid to total flux ratio as a
  function of magnitude for all the galaxies in a 1~Mpc radius within
  clusters at redshift $0.4 \leq z < 0.75$. }
\label{fig:errratio_mabs}
\end{figure}

Before stacking clusters and searching for variations of galaxy
morphologies with redshift, it is necessary to make a cut in absolute
magnitude in order to have comparable samples in all the redshift
bins. We make the choice of the limiting magnitude by considering
  the redshift range that imposes the strongest constraints on the
  relative 
  uncertainty on the spheroid-to-total ratio: $0.4 \leq z < 0.75$.  A
  plot of this uncertainty as a function of absolute magnitude for all
  the cluster galaxies in the redshift range $0.4 \leq z < 0.75$
  is shown in Fig.~\ref{fig:errratio_mabs}. We choose to limit the
  relative uncertainty on the spheroid to total flux ratio to 20\% and
  to cut the sample at ${\rm M_r \leq -19.0}$ (which roughly
  corresponds to $M^*+3$).  Out of the initial sample of 1,574,505
  galaxies, there are 1,128,389 galaxies with ${\rm M_r \leq -19.0}$,
   of which 522,605 have an uncertainty $err_{flux ratio}$ on the
  spheroid-to-total flux ratio smaller than or equal to 20\%.  So for
  $M_r\le -19.0$ we can consider that about 50\% of the galaxies have
  $err_{flux ratio} \leq$20\%. Hereafter we will take into account
  only the galaxies with an absolute magnitude brighter than ${\rm M_r
    < -19.0}$.

In the following analysis, we will limit our analysis to the 2530
clusters detected at a 4$\sigma$ level and above to have a sample of
clusters that is as reliable as possible. 
We stacked clusters in six redshift bins: $z\leq 0.15$, $0.15<z\leq
0.25$, $0.25<z\leq 0.35$, $0.35<z\leq 0.45$, $0.45<z\leq 0.55$, and
$z>0.55$ and computed the percentages of late-type galaxies.  If we
assume that there is no observational bias due to the loss of spatial
resolution for galaxies when redshift increases,  we find that the
percentage of late-type galaxies tends to decrease with redshift,
opposite to what is expected. We also stacked clusters in four bins
of detection level: 4$\sigma$, 5$\sigma$, 6$\sigma$, and 9$\sigma$,
which roughly correspond to cluster mass bins. Here too, we tend to
find that the percentage of late-type galaxies somewhat increases with
significance level, the opposite of what is expected (more massive
clusters are expected to host more early-type galaxies).
We therefore performed simulations to test the hypothesis
that these unexpected results could be due to an observational bias.

\subsection{Influence of the redshift on the morphological classification}

\begin{figure}[h!]
\centering
  \resizebox{\hsize}{!}{\includegraphics[angle=-90, viewport=50  50 600 850,clip]{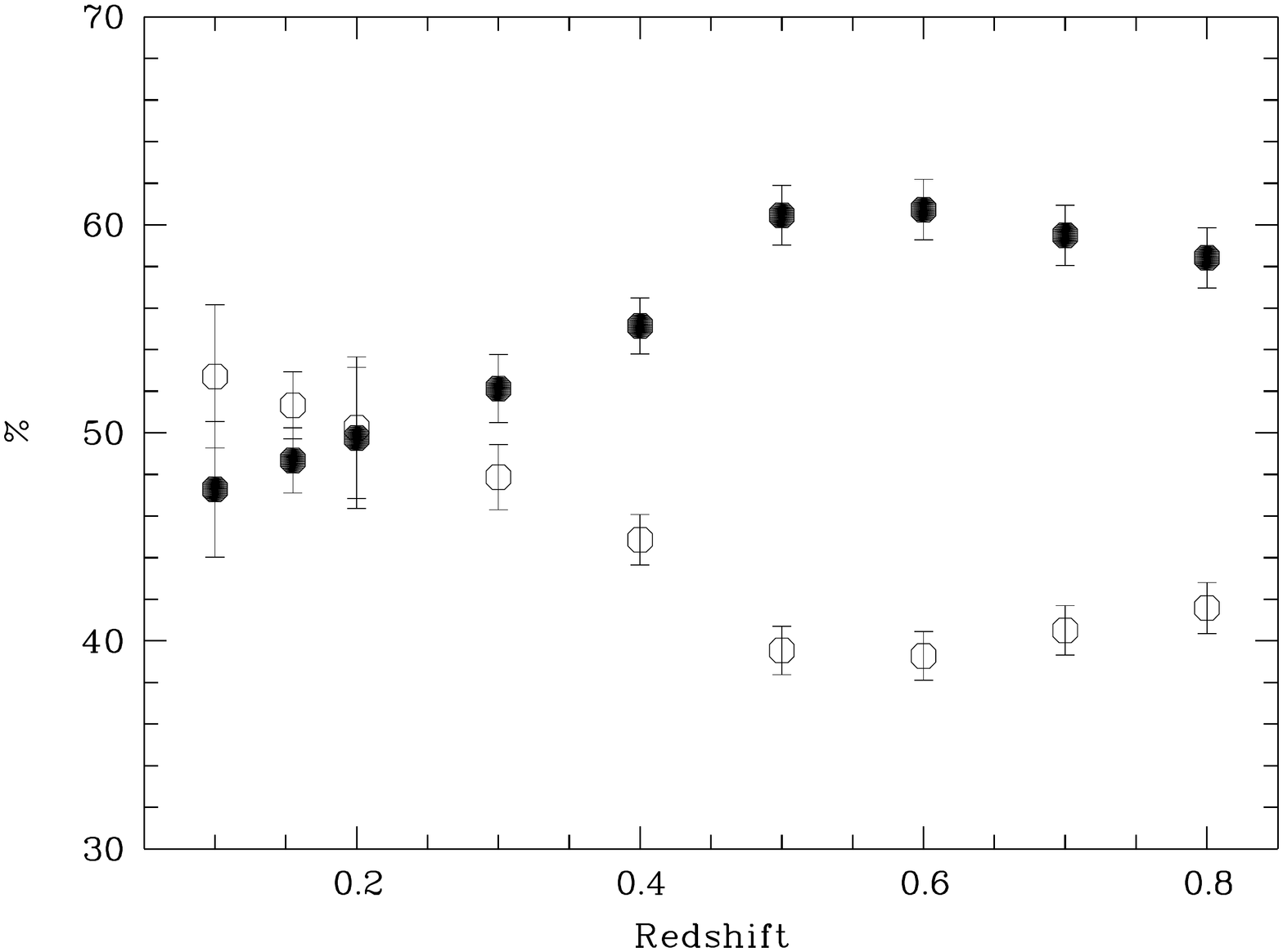}}
  \caption[]{Percentage of early-type galaxies (filled black circles) and late-type galaxies (empty black circles) as a function of redshift obtained by
  stacking 103 clusters artificially degraded to mimic the influence
  of increasing redshift (see Section~6.2).}
\label{fig:simu}
\end{figure}

In order to test how the image degradation due to increasing redshift
could influence the value of the
$f_{spheroid}/(f_{disk}+f_{spheroid})$ flux ratio on which our late-
and early-type galaxy percentages are based, we selected 103 clusters
with redshift $z \le 0.20$ and detected at least at the $4\sigma$
level.  Starting from the original images, we artificially degraded
the images by rebinning them to larger pixel sizes to mimic the effect
of increasing redshift. In this way images were computed to simulate
the clusters as if they were located at redshifts between 0.2 and 0.8,
in bins of 0.1 in redshift. The rebinned images were then treated with
SExtractor as above to compute the
$f_{spheroid}/(f_{disk}+f_{spheroid})$ flux ratios of all the cluster
galaxies.

The percentages of late- and early-type galaxies were then stacked in
redshift bins, and the results are shown in Fig.~\ref{fig:simu}.  This
figure clearly shows that, as a bias due to redshift, the percentage
of late-type galaxies tends to decrease with redshift and that of
early types to increase.  Therefore, when estimating the early-to-late-type ratio, a correcting factor must be applied to correct for
this bias.  The number of late-type galaxies for various redshifts is
given in Table~\ref{tab:corr}. We  note that we  only
consider here cluster galaxies, for which the computed
absolute magnitudes take into account the k-corrections and luminosity
distance corrections.

\begin{table}
\centering
\caption{Percentage of late-type galaxies as a function of redshift
  obtained by degrading the images  to mimic the effect of
  increasing redshift, as explained in the text.}
\begin{tabular}{ccc}
\hline
\hline
Redshift & \% of late types & number of galaxies\\
\hline
0.1    & 52.7  &  4005 \\
0.155  & 51.3  & 17824 \\
0.2    & 50.2  &  3893 \\
0.3    & 47.9  & 17543 \\
0.4    & 44.9  & 27283 \\
0.5    & 39.5  & 26191 \\
0.6    & 39.3  & 25895 \\
0.7    & 40.5  & 25652 \\
0.8    & 41.6  & 25003 \\
\hline
\end{tabular}
\label{tab:corr}
\end{table}

\subsection{Results}

\begin{figure}[h!]
\centering
  \resizebox{\hsize}{!}{\includegraphics[viewport=15  15 750 500,clip]{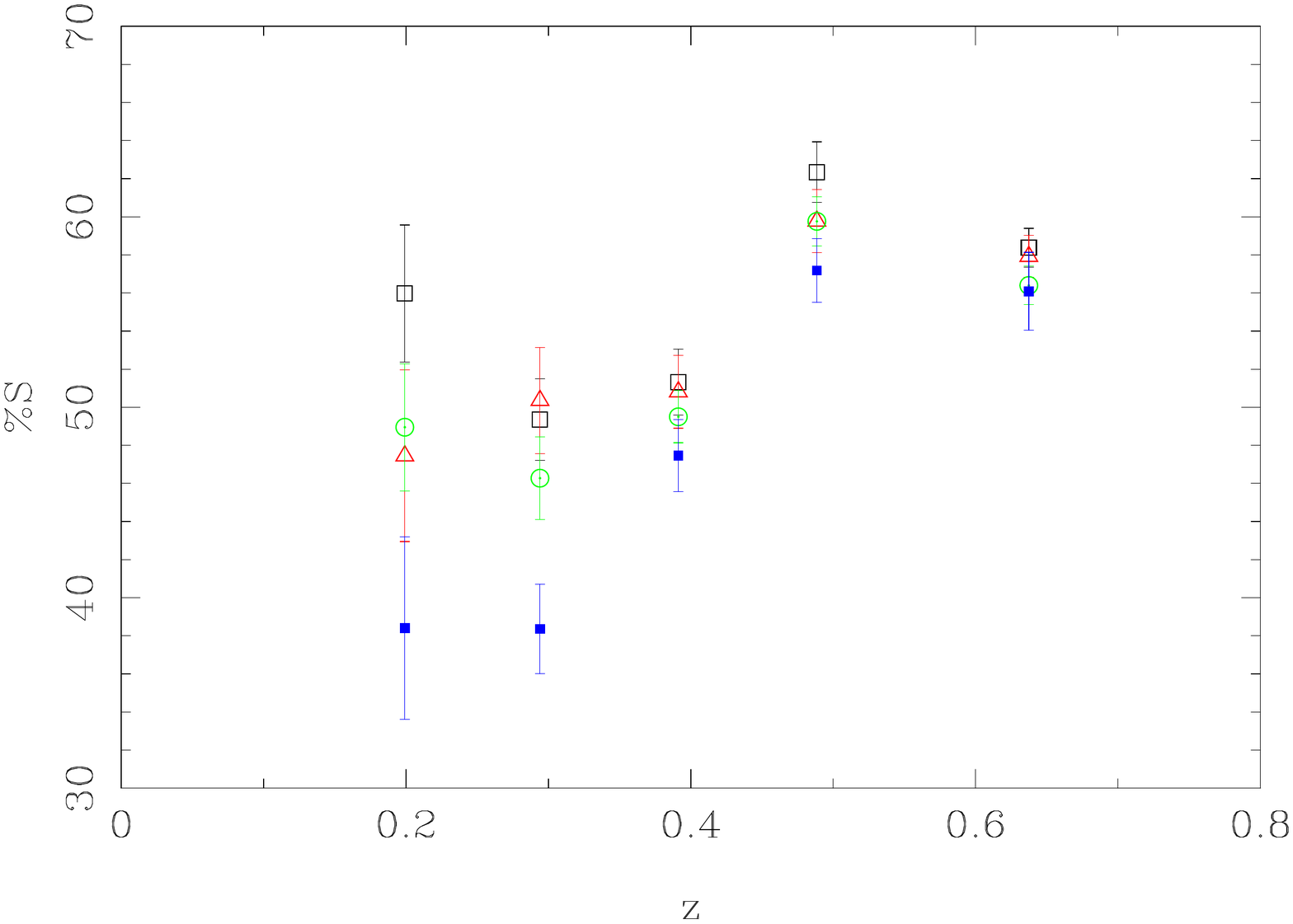}}
  \caption[]{Percentage of late-type galaxies as a function of redshift,
  based on the bulge to disk decomposition in the $r'$ band. The data
  points are colour--coded as a function of detection level: black
  circles for 4$\sigma$, red triangles for 5$\sigma$, green circles for
  6$\sigma$, and blue squares for 9$\sigma$. The correction factors
  explained in the text have been applied. }
\label{fig:SE_z}
\end{figure}

\begin{figure}[h!]
\centering
  \resizebox{\hsize}{!}{\includegraphics[viewport=15  15 750 500,clip]{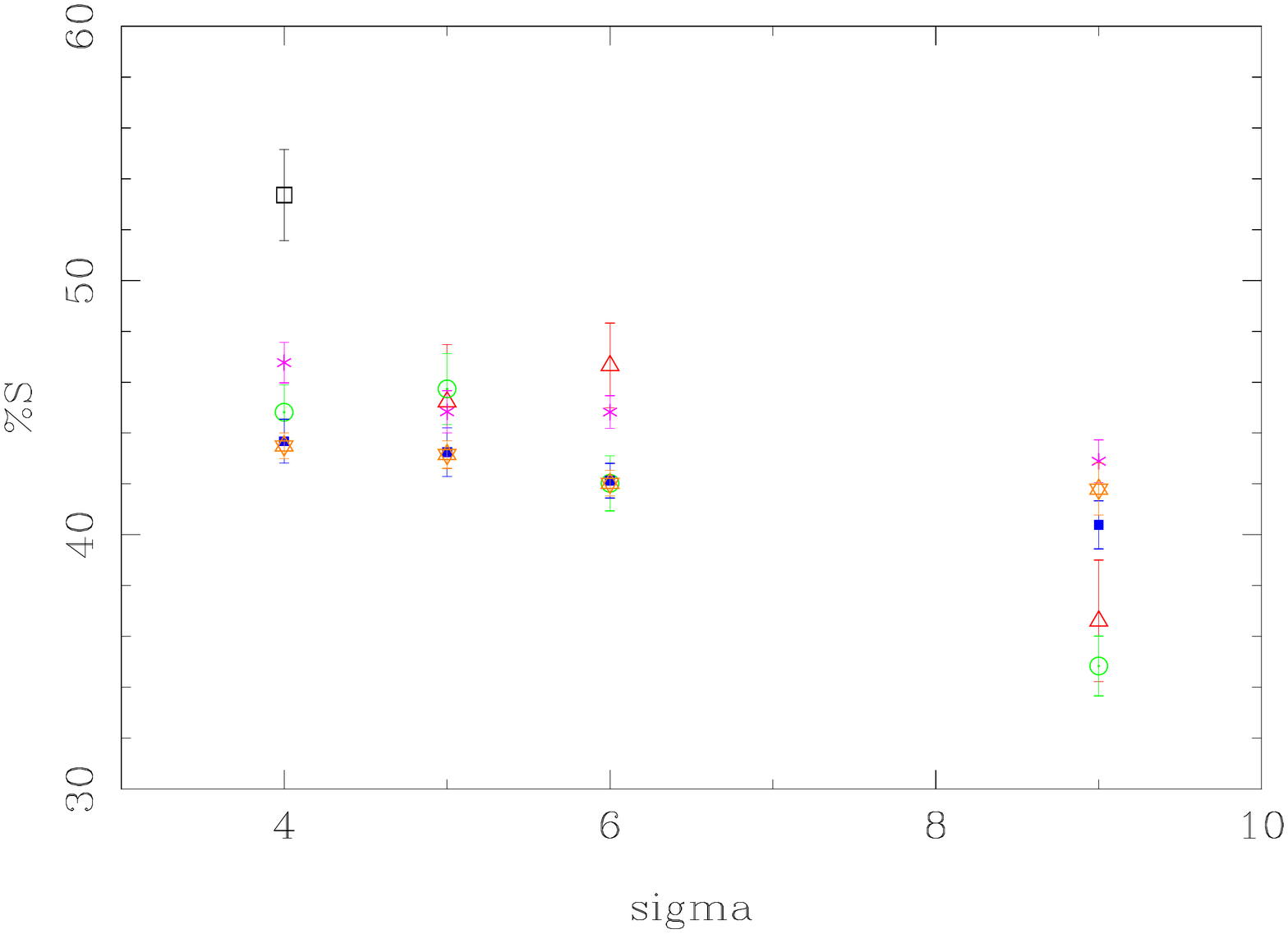}}
  \caption[]{Percentage of late-type galaxies as a function of detection
  level, based on the bulge to disk decomposition in the $r'$ band. The
  data points are colour--coded in bins of redshift: black squares for
  $z\leq 0.2$, red triangles for $0.2<z \leq 0.27$, green circles for
  $0.27<z\leq 0.37$, blue squares for $0.37<z\leq 0.47$, magenta
  crosses for $0.47<z\leq 0.57$, and orange stars for $z>0.57$. The
  correction factors explained in the text have been applied. }
\label{fig:SE_sigma}
\end{figure}

If we apply the correction factors derived from Table~\ref{tab:corr} to
the percentages of late-type and early-type cluster galaxies
found above, we obtain the results displayed in Figs.~\ref{fig:SE_z}
and \ref{fig:SE_sigma}. In these two figures, the error bars were taken to
be Poissonian: $\sqrt N/N$, where N is the number of early-type
galaxies corresponding to each point.

We can see that the percentages of late-type galaxies increase with
redshift. This is particularly visible for 9$\sigma$ clusters, where
the percentage of late types increases from 20\% to almost 60\%
between redshifts z=0.2 and z=0.5.

The percentages of late-type galaxies show a trend of decreasing with
detection level (i.e. with cluster mass). We  note that the
percentages of late-type galaxies that we find are notably higher than
those of Postman et al. (2005) or Smith et al. (2005) perhaps because  our classification of late- and early-type galaxies
is not the same, and/or because our cluster galaxies
are probably at least partly contaminated by field galaxies.

\subsection{Comparison of the galaxy type classifications 
by SExtractor and Le~Phare}

\begin{figure}
\centering
  \resizebox{\hsize}{!}{\includegraphics[viewport=0 15 770 470,clip]{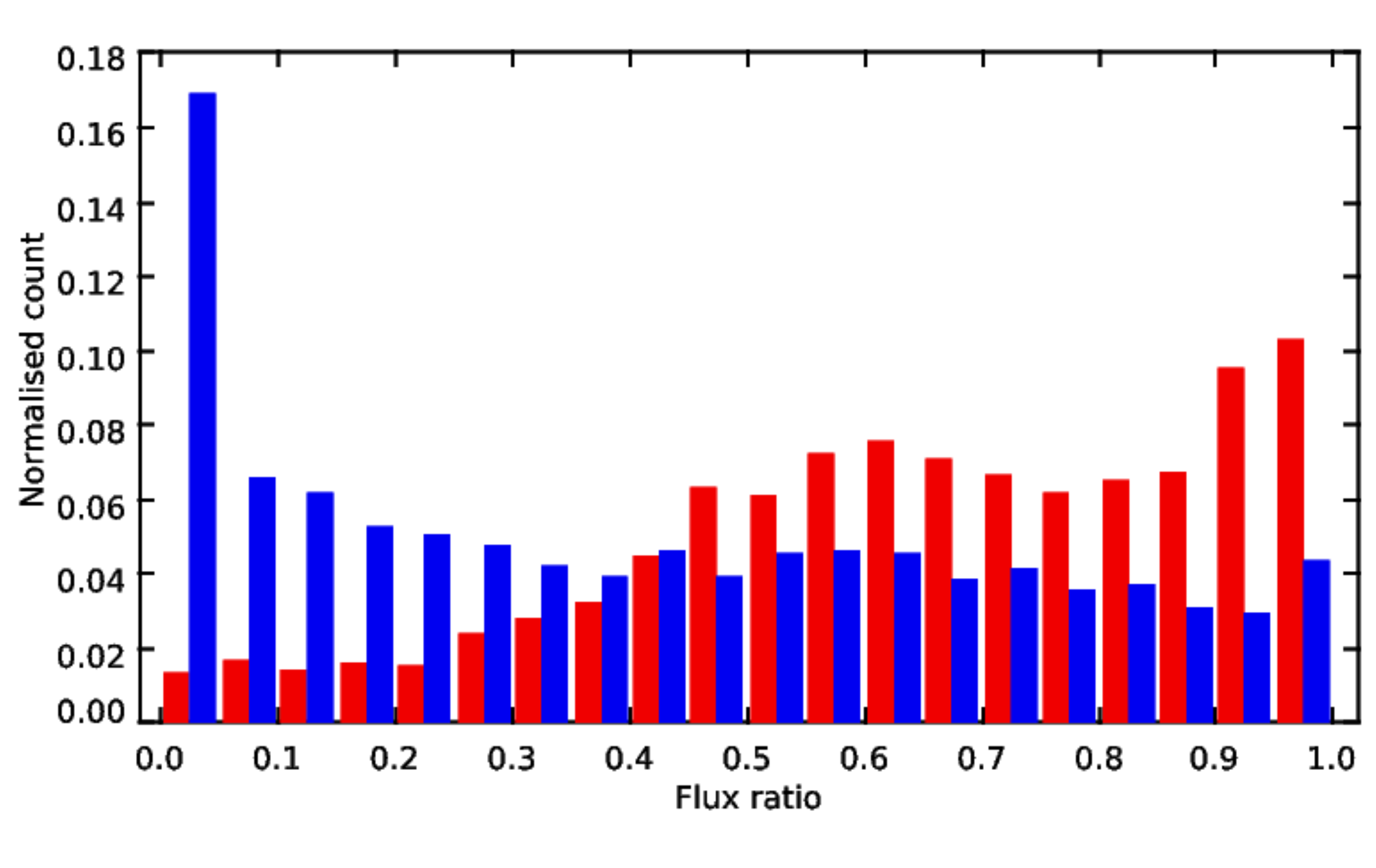}}
  \caption[]{Histograms of the normalized counts of late-type (in blue)
and early-type (in red) galaxies as classified by Le~Phare,
as a function of spheroid-to-disk flux ratio.}
\label{fig:classif}
\end{figure}

Since we had to run the Le~Phare software to compute the absolute
magnitudes of the cluster galaxies in order to calculate GLFs,  as a by-product we obtained a Le~Phare galaxy type classification (the
same one as that used in the COSMOS survey).  Le~Phare assigns each
galaxy a type coded as a number between 1 and 31, with early-type
galaxies between 1 and 7, late-type galaxies between 8 and 19, and AGN
between 20 and 31. These types correspond to the best spectral
template allowing a  fit to the photometric data.

The early- and late-type classifications that we made with SExtractor
based on pure morphological properties are not expected to match
exactly those derived with Le~Phare. However, we believe it is
interesting to compare them on a large statistical basis.

If we take into account all the cluster galaxies (77,162
galaxies), we find that 70\% of the early-type and 53\% of the late-type galaxies have the same classification with the two methods, after
eliminating the AGN and starburst types from Le~Phare (which add noise
to the final morphological classification).

We cross-identified the cluster galaxies with the spectroscopic
catalogue described in Section~2.1.  The sample is then
reduced to only 8,105 galaxies.  For this sample, we find that 74\% of
the early-type galaxies are well classified by both methods, 61\% of the
late types and 68\% if we add late types and AGN.

As a test, we also considered 73,970 galaxies in the  Stripe~82
region having a spectroscopic redshift available, independent of  whether they were 
cluster galaxies or not. We ran Le~Phare on
those galaxies, fixing their photo$-z$ to be equal to their
spectroscopic redshift to obtain the best possible Le~Phare type.
We find that  69\% of the early-type and 58\% of the late-type galaxies have the same classification with Le~Phare and SExtractor.
This percentage becomes 63\% if we add late types and AGN. 

As an illustration, we show in Fig.~\ref{fig:classif} the histograms
of the normalized counts of late- and early-type galaxies as classified
by Le~Phare as a function of spheroid-to-disk flux ratio computed by
SExtractor. 

Since these two ways of classifying galaxies are very different from
one another (one being purely morphological while the other is purely
spectral), and since morphological and spectral evolutions can also be
quite different, it is rather satisfying to see that they agree
between 58\% and 74\% of the cases.

\subsection{Eye-test of the morphological classification}

In order to test the morphological classification obtained with
SExtractor, six high school students (see their names in the
acknowledgements) selected about 1000 galaxies in the redshift range
$0.15<z<0.25$ classified as early-type or late-type and examined  them
visually with ds9 one by one.  They found that the SExtractor and eye
classifications agreed for $80\pm$10\% of the galaxies.

\section{Summary and conclusions}
\label{sec:concl}

Based on the galaxy photometric redshift catalogue of Reis et
al. (2012), we have searched for galaxy clusters in the Stripe~82
region of the Sloan Digital Sky Survey by applying the AMACFI cluster
finder (Adami \& Mazure 1999). After making nine tests with different
AMACFI parameters that have a strong influence on the cluster
detection rate, we detected 3663 candidate clusters at a 3$\sigma$
level and above, in the redshift range $0.1\leq z \leq 0.7$, with
estimated mean masses between $\sim 10^{13}$ and a few 10$^{14}$
M$_\odot$.  We cross-correlated our catalogue of candidate clusters
with various catalogues extracted from optical and/or X-ray data.  The
percentages of redetected clusters are at most 40\%, but in all cases
this can be explained by the fact that we detect relatively massive
clusters, while other authors detect less massive structures.

The colour-magnitude diagrams and galaxy luminosity functions of the
clusters detected at 5$\sigma$ and above and stacked in redshift bins
of width 0.1 are typically those of {\it bona fide} clusters. This
confirms that the clusters we have detected have  a high
probability of being real clusters.

The morphological analysis of the cluster galaxies shows that
the fraction of late-type to early-type galaxies shows an increase
with redshift and a decrease with significance level, i.e. cluster
mass. This result is obtained after correcting for a bias due to the
effect of increasing redshift that we quantified through simulations.

Although the 3663 candidate clusters detected here seem mostly to
be real clusters, spectroscopic confirmation would of course be
necessary. We are in the process of improving the positions and
redshifts of our clusters by searching for the brightest cluster
galaxies, and retrieving spectroscopic redshifts in the SDSS data base.
As yet another confirmation to the reality of the clusters detected in
S82, we are also identifying our candidate clusters with diffuse X-ray
sources detected by XMM-Newton when available.
These results will be published in a forthcoming paper.

Counting the number of clusters per unit volume and the growth of
clusters with redshift are methods for delimiting cosmological model
parameters such as $w$,  $dw/dz$, and $\sigma_8$ (Allen et
al. 2011). This motivated the present search for clusters in the
Stripe~82 region of the SDSS, as well as our previous searches for
clusters in the CFHTLS. In the near future, the Dark Energy Survey
expects to find approximately 170,000 clusters with masses $\geq 5
\times 10^{13}$~M$_\odot$
(http://en.wikipedia.org/wiki/The\_Dark\_Energy\_Survey), and LSST
more than 100,000 clusters with masses $\geq 2\times 10^{14}$~M$_\odot$
(Tyson et al. 2003).

Based on our experience here, we conclude that is it very important
not to depend on using just one cluster detection algorithm.
Therefore,  for future surveys we suggest the following approach to
derive cosmological parameters from optical/near IR cluster surveys:
1)~take a $ \sim 6\sigma$ cut and a $\sim 4\sigma$ cut; and 2)~estimate
the completeness of the survey by comparing two or more different
cluster finding techniques.  The derived cosmological parameters based
on two (or more) different $\sigma$ cuts and techniques can then be
used to determine the underlying systematic limits to the values of
these cosmological parameters.

\begin{acknowledgements}

  F.D. acknowledges long-term support from CNES. I.M. acknowledges
  financial support from the Spanish grants AYA2010-15169 and
  AYA2013-42227-P and from the Junta de Andalucia through TIC-114 and
  the Excellence Project P08-TIC-03531. A.T. acknowledges the support
  and the hospitality of IAP/CNRS for two one--month visits.

  We are grateful to Andrea Biviano for giving us his IDL program to
  fit GLFs with a Schechter function and to Alberto Cappi for
  discussions. We thank the six high school students M.A.~Garc\'\i a
  Valverde, B.~Hern\'andez Ramos, J.~Leon Lovell, L.~Mart\'\i nez
  S\'anchez de Lara, J.~Rodr\'\i guez Zamorano and L.~Vallecillos Azor
  for their careful eye classification of about 1000 galaxies.

  Funding for SDSS-III has been provided by the Alfred P. Sloan
  Foundation, the Participating Institutions, the National Science
  Foundation, and the U.S. Department of Energy Office of Science. The
  SDSS-III web site is \url{http://www.sdss3.org/}.

SDSS-III is managed by the Astrophysical Research Consortium for the
Participating Institutions of the SDSS-III Collaboration including the
University of Arizona, the Brazilian Participation Group, Brookhaven
National Laboratory, Carnegie Mellon University, University of
Florida, the French Participation Group, the German Participation
Group, Harvard University, the Instituto de Astrofisica de Canarias,
the Michigan State/Notre Dame/JINA Participation Group, Johns Hopkins
University, Lawrence Berkeley National Laboratory, Max Planck
Institute for Astrophysics, Max Planck Institute for Extraterrestrial
Physics, New Mexico State University, New York University, Ohio State
University, Pennsylvania State University, University of Portsmouth,
Princeton University, the Spanish Participation Group, University of
Tokyo, University of Utah, Vanderbilt University, University of
Virginia, University of Washington, and Yale University.

\end{acknowledgements}

\appendix

\section{Magnitude histograms}

\begin{figure*}[t!]
\centering
  \resizebox{5cm}{!}{\includegraphics[viewport=15  150 590 710,clip]{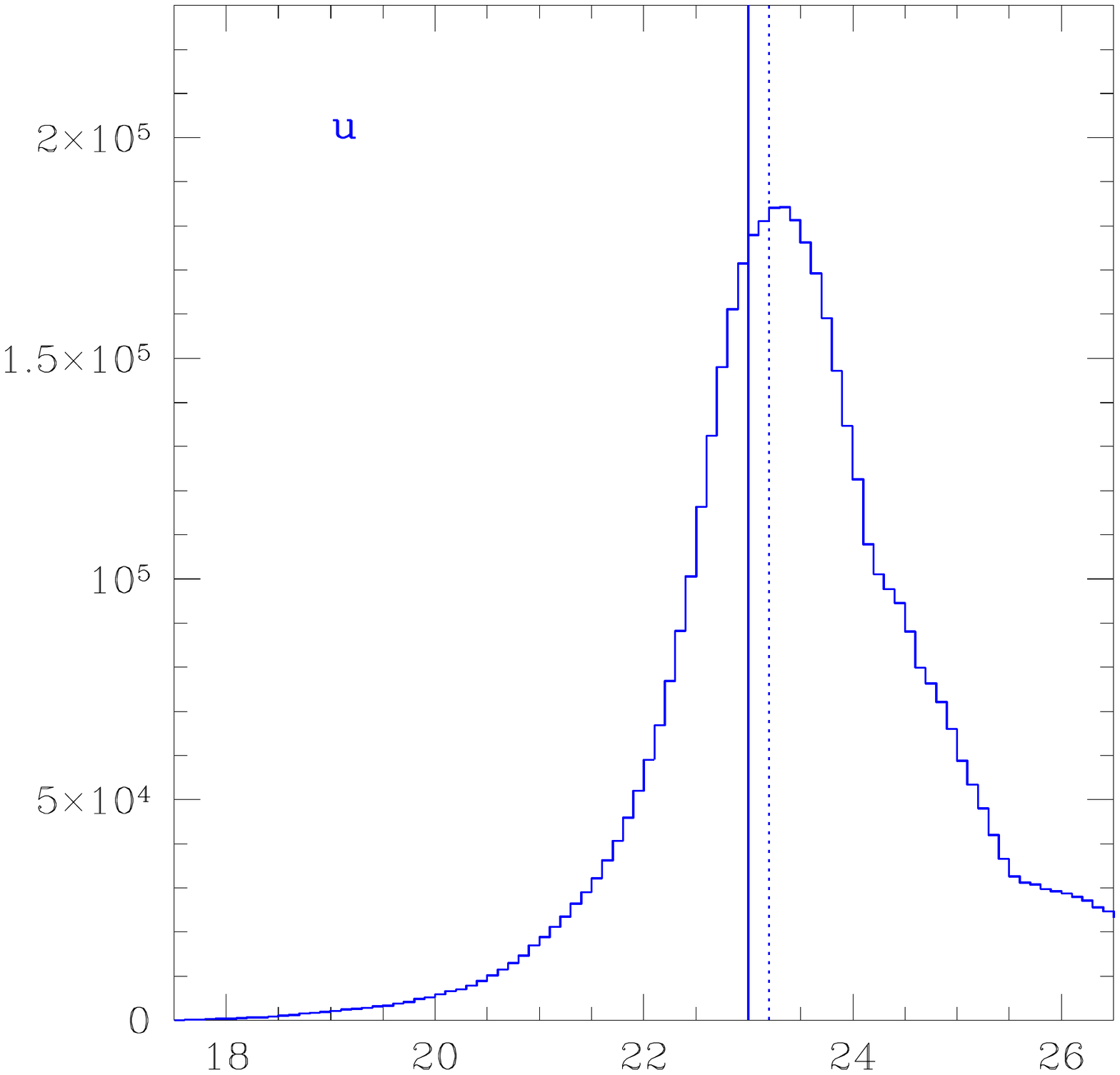}}
  \resizebox{5cm}{!}{\includegraphics[viewport=15  150 590 710,clip]{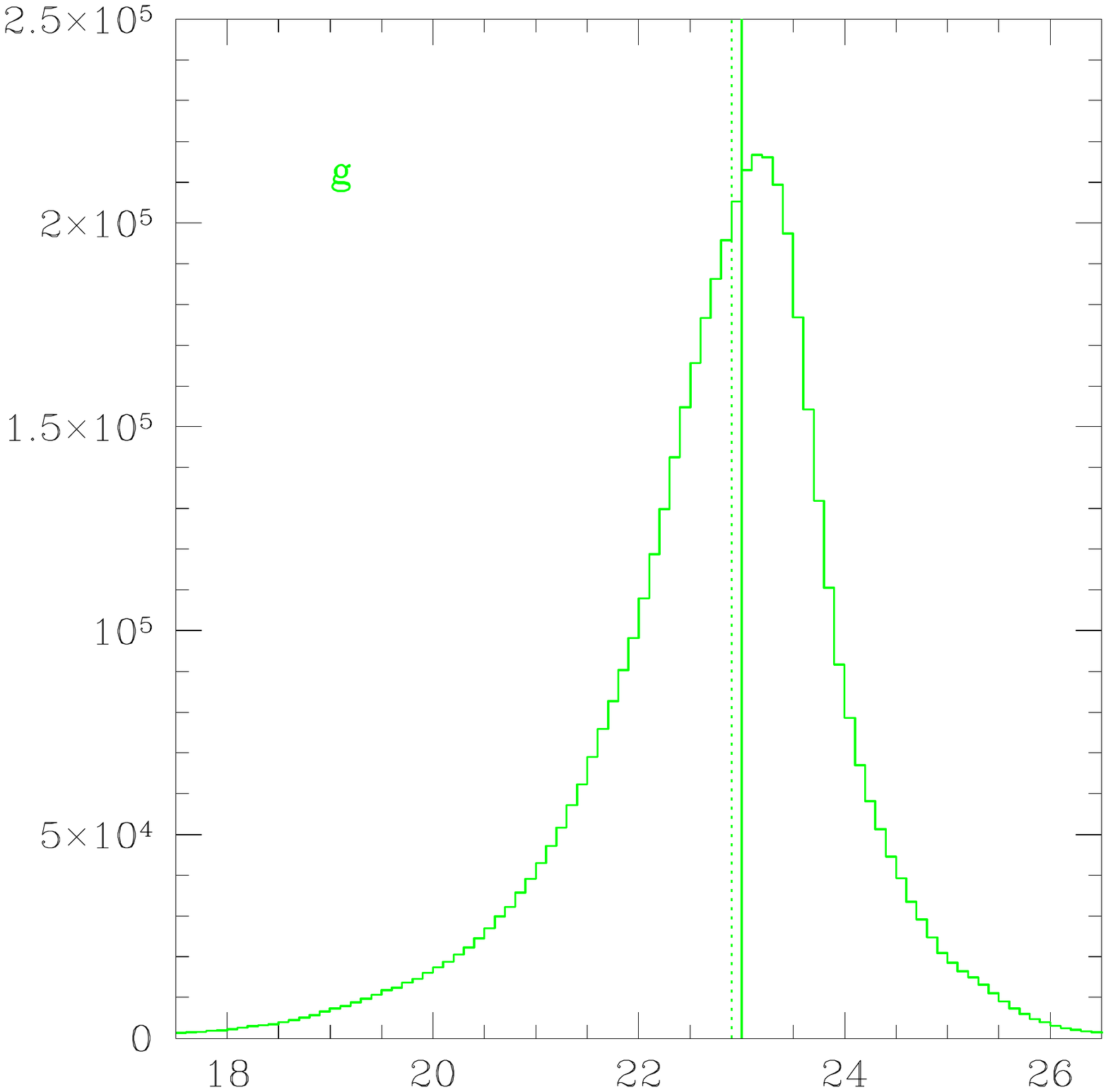}}
  \resizebox{5cm}{!}{\includegraphics[viewport=15  150 590 710,clip]{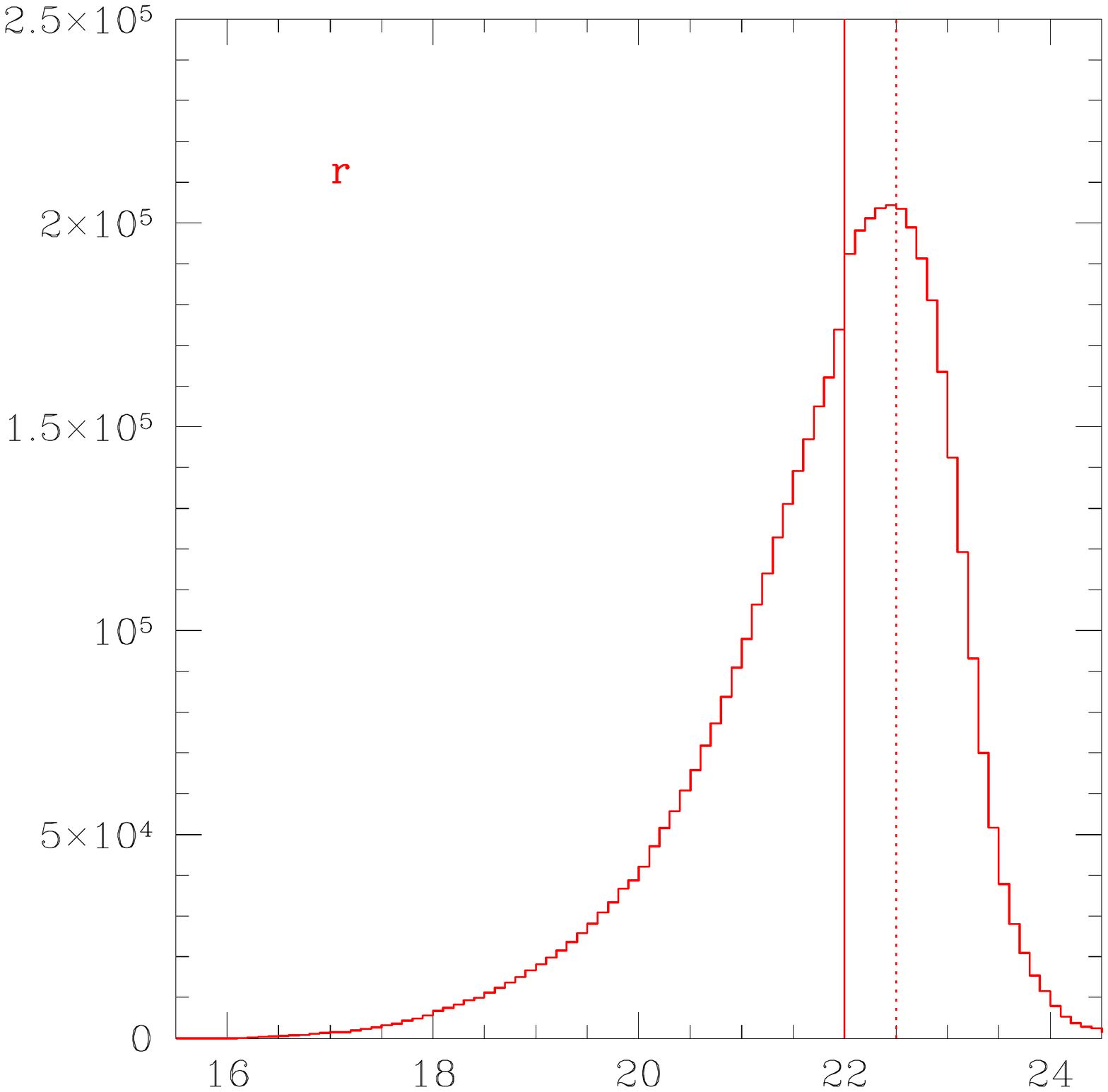}}
  \resizebox{5cm}{!}{\includegraphics[viewport=15  150 590 710,clip]{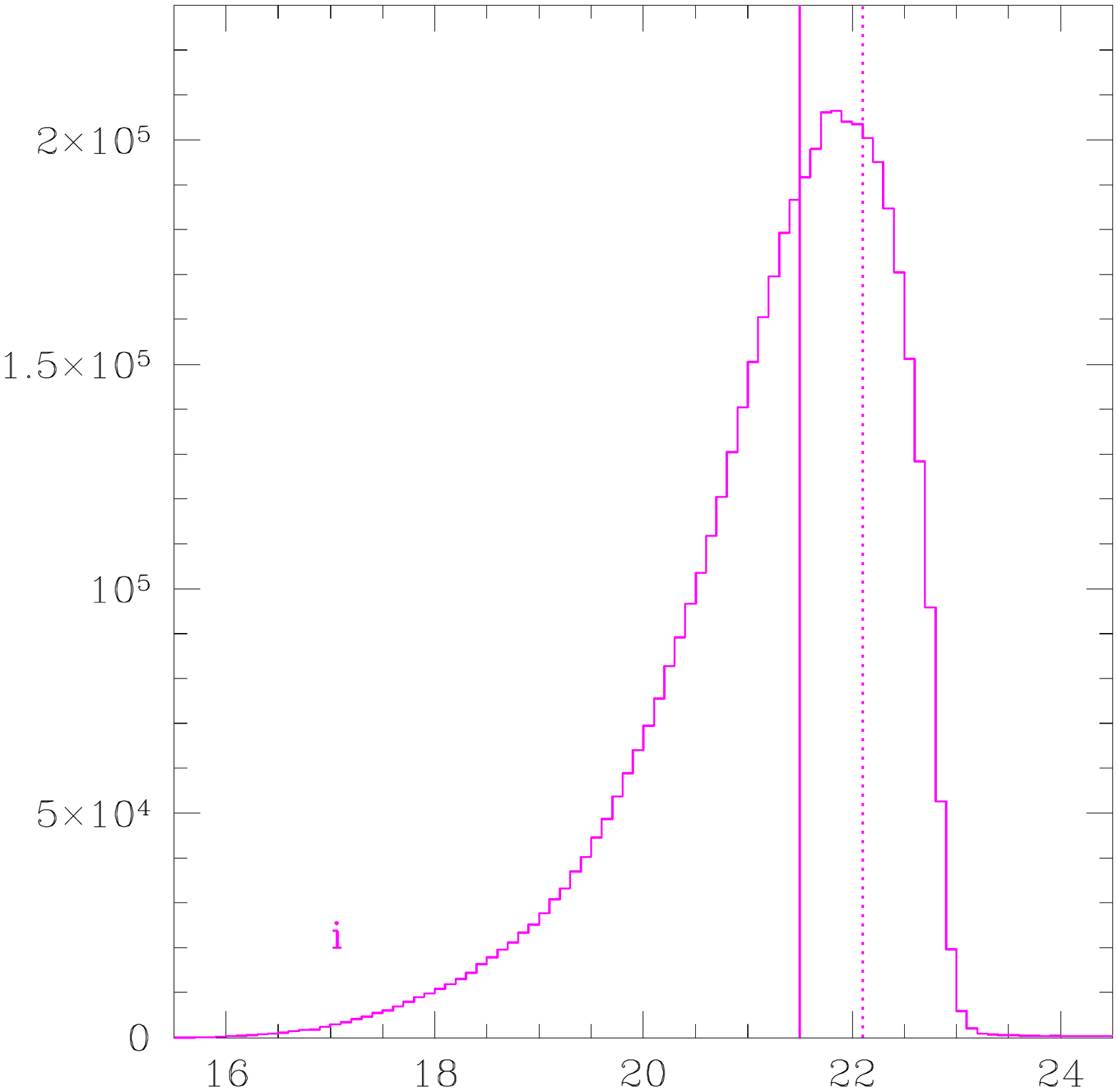}}
  \resizebox{5cm}{!}{\includegraphics[viewport=15  150 590 710,clip]{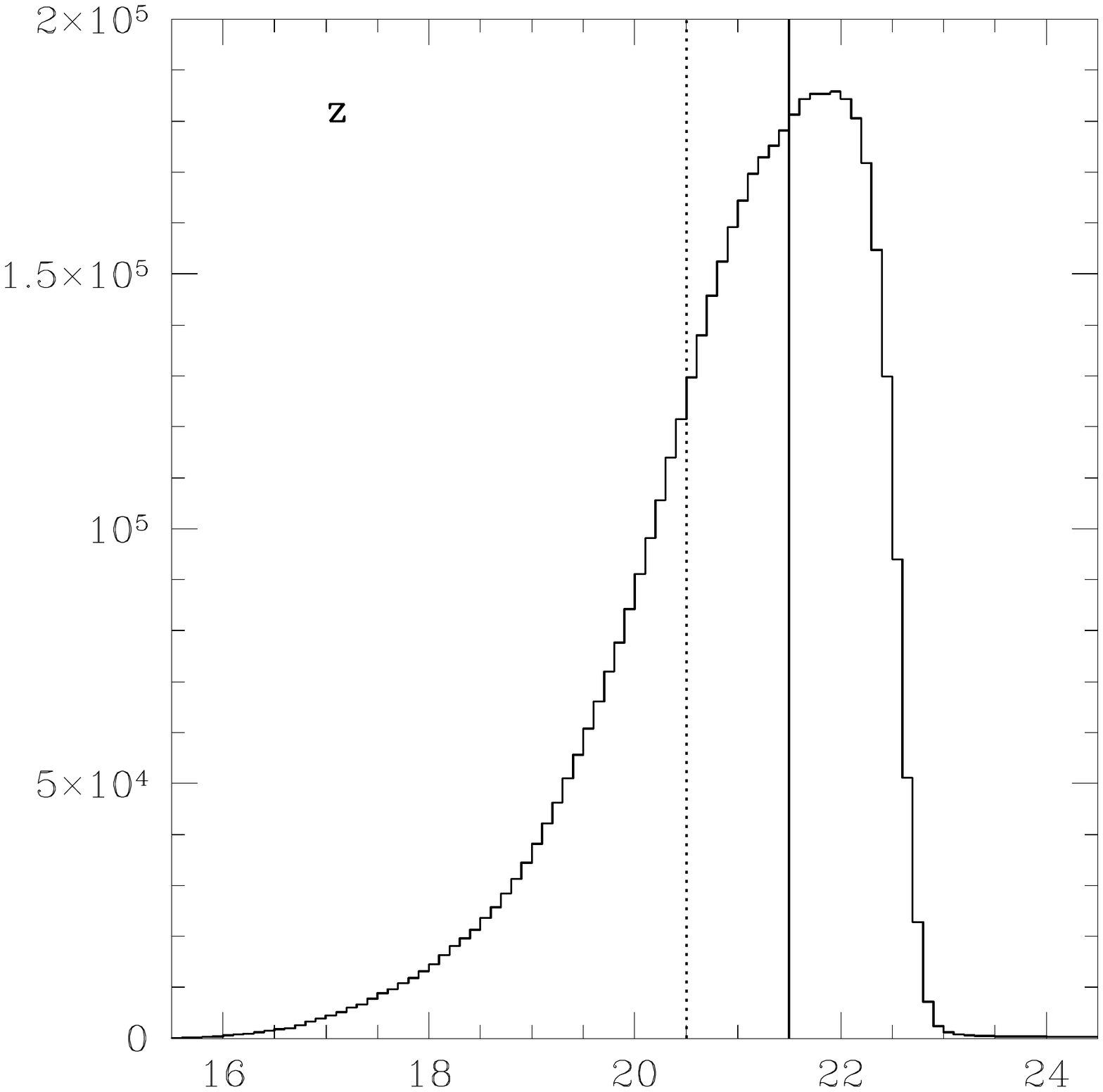}}
  \caption[]{Magnitude histograms in the five bands of the 4,999,968
  galaxies of the initial magnitude catalogue used to compute absolute
  magnitudes. The full vertical lines show the 90\% completeness
  limits beyond which the galaxy counts will not be considered as fitting
  the GLFs. These limits are: $u'_{lim}$=23.0, $g'_{lim}$=22.8,
  $r'_{lim}$=22.1, $i'_{lim}$=21.5, and $z'_{lim}$=21.2. The dotted
  vertical lines show the 90\% completeness limits derived from Fig.~8
  in Annis et al. (2014) for comparison.  }
\label{fig:histomagall}
\end{figure*}

The magnitude histograms in the five bands of the 4,999,968 galaxies
of the initial magnitude catalogue used to compute absolute magnitudes
are shown in Fig.~\ref{fig:histomagall}.

\section{Completeness simulations}

\begin{figure}[h!]
\centering
  \resizebox{\hsize}{!}{\includegraphics[angle=-90,viewport=20 20 600 850,clip]{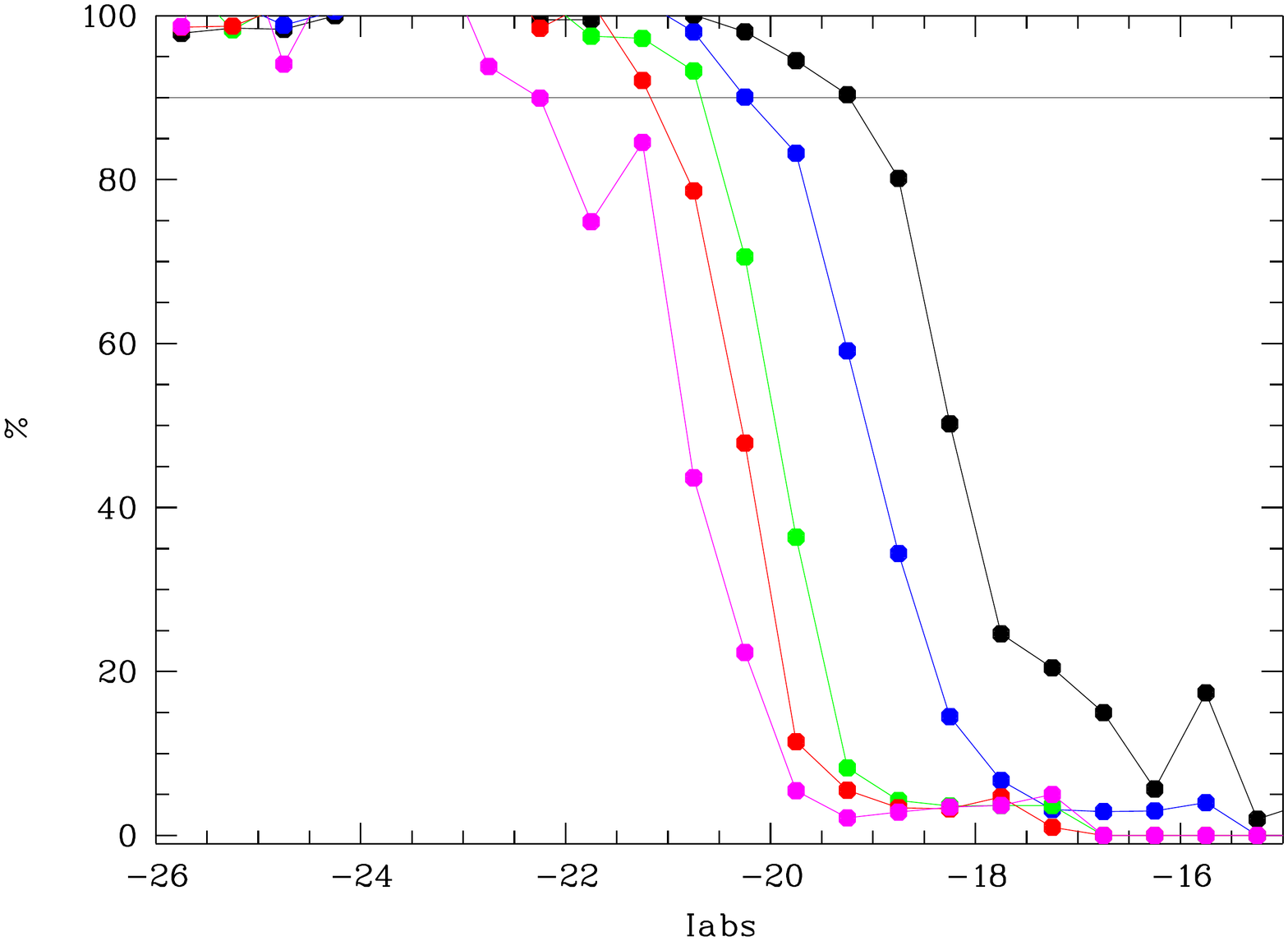}}
  \caption[]{Percentage of redetected galaxies as a function of absolute
  magnitude in the $i'$ band derived from our simulations in five
  magnitude bins: z=0.2 in black, z=0.3 in blue, z=0.4 in green, z=0.5
  in red, and z=0.6 in magenta (see Section 5.2.1).}
\label{fig:simucompl}
\end{figure}

We show in Fig.~\ref{fig:simucompl} the percentages of redetected
galaxies as a function of absolute magnitude in the $i'$ band derived
from our simulations in five magnitude bins: z=0.2 in black, z=0.3 in
blue, z=0.4 in green, z=0.5 in red, and z=0.6 in magenta (see Section
5.2.1). Simulations in the other bands give comparable curves and are
not shown here to save space.

\end{document}